\documentclass[12pt]{iopart}

\usepackage{iopams} 
\usepackage{setstack} 
\usepackage{graphicx}
\usepackage{subfig}
\usepackage{caption}
\usepackage[dvips]{epsfig}
\usepackage[usenames]{color}
\usepackage[normalem]{ulem}

\newcommand{\nmax}{n_{\mathrm{max}}} 
\newcommand{\peq}{p^{\mathrm{eq}}_n(\mu)}
\newcommand{\nav}{\langle n \rangle}
\newcommand{\nvar}{\langle n^2 \rangle - \langle n \rangle^2}

\begin{document}

\title[Current fluctuations in boundary driven diffusive systems]{Current fluctuations in boundary driven diffusive systems in different dimensions: a numerical study}

\author{T Becker$^1$, K Nelissen$^{2,1}$, B Cleuren$^1$}
\address{$^1$ Hasselt University, B-3590 Diepenbeek, Belgium}
\address{$^2$ Departement Fysica, Universiteit Antwerpen, Groenenborgerlaan 171, B-2020 Antwerpen, Belgium}
\ead{thijsbecker@gmail.com}

\vspace{10pt}
\begin{indented}
\item[]\today
\end{indented}

\begin{abstract}
We use kinetic Monte Carlo simulations to investigate current fluctuations in boundary driven generalized exclusion processes, in different dimensions. Simulation results are in full agreement with predictions based on the additivity principle and the macroscopic fluctuation theory. The current statistics are independent of the shape of the contacts with the reservoirs, provided they are macroscopic in size. In general, the current distribution depends on the spatial dimension. For the special cases of the symmetric simple exclusion process and the zero-range process, the current statistics are the same for all spatial dimensions.
\end{abstract}

%
\vspace{2pc}
\maketitle
%
%

\section{Introduction}

A system connected to two particle reservoirs at different densities relaxes to a nonequilibrium steady state (NSS), with a particle current flowing through it. The description of the fluctuations of this current has recently received much attention \cite{PREAppert,JSP_Hurtado,hurtado2009current,PNAShurtado,prolhac2009cumulants,PRE_mieke_2009,gorissen2011finite,PRL_mieke_2012,kundu2011large,lazarescu2011exact,maes2008steady,villavicencio2012,polettini2014transient,JSM_vilenking,villavicencio2014fluctuation}. In equilibrium, thermodynamic potentials are related to exponentially unlikely fluctuations away from the average \cite{Touchette20091}, as was already discussed by Einstein \cite{Einstein1910}. 
Analogously, one can construct nonequilibrium thermodynamic potentials from the study of exponentially unlikely current and density fluctuations away from the NSS \cite{JSM_jona_2014}. A theoretical framework for this approach is provided by the macroscopic fluctuation theory (MFT) \cite{PRL_Bertini_2001,bertini2002macroscopic,MFT_PRL_2005,MFT_JSP_2006,bertini2014macroscopic}. 

Using the MFT, Akkermans and co-workers studied current fluctuations in diffusive systems connected to two reservoirs \cite{EPL2013}. They showed analytically that for a system of arbitrary (but fixed) dimension, the ratio of the cumulants of the current distribution is independent of the shape of the system and the shape of the contacts with the reservoirs. This derivation is valid if both the system and the contacts with the reservoirs are macroscopic in size. The analytical prediction was tested by numerically calculating the ratio of the first two cumulants, called the Fano factor, for the symmetric simple exclusion process (SSEP). In two dimensions, convergence to the analytical predictions was found for large system sizes by assuming a power-law behavior and extrapolating the numerical data. In three dimensions no convergence was found. The numerical results were, however, obtained for contacts that are not macroscopic in size. Akkermans et al.~therefore argued that the discrepancy between numerics and theory was caused by too small contact sizes with the reservoirs.

Under certain conditions, the asymptotic current distribution of a one-dimensional system that is described by the MFT can be calculated from an additivity principle (AP) postulated by Bodineau and Derrida \cite{bodineau2004}. The validity of this AP has been confirmed in several one-dimensional systems, both analytically  \cite{bodineau2004,bodineau2007cumulants,van2005large,harris2005current,PRE_marko} and numerically \cite{JSP_Hurtado,PRE_marko,PRLhurtado,PRE_Hurtado_AP,PRE_mieke_2012}. An interesting question is if one can use the AP to predict the current distribution in higher-dimensional systems. This is especially important because many experimental systems are higher-dimensional. The results from \cite{EPL2013} indicate that it is, indeed, possible to do this. So far, only a few studies have addressed this question. Saito and Dhar studied heat fluctuations in a deterministic system connected to stochastic reservoirs \cite{PRLSaito}. They found that the AP can predict the current distribution in three dimensions, both for diffusive and anomalous heat transport. Hurtado, P{\'e}rez-Espigares, del Pozo, and Garrido confirmed the validity of the AP for the two-dimensional Kipnis-Marchioro-Presutti (KMP) model \cite{JSP_Hurtado,pe2011large_article}.

We study numerically the first three cumulants of the current distribution of boundary driven generalized exclusion processes (GEPs) \cite{kipnis}. The dynamics is simulated using kinetic Monte Carlo (kMC). The simplest case of a GEP is the SSEP, where only one particle can occupy each lattice site. In our simulations of the SSEP we consider contacts with the reservoirs that are macroscopic in size. Complete convergence of the Fano factor to the analytical prediction of \cite{EPL2013} is found in two dimensions. For three dimensions the data indicate convergence for large system sizes. We proceed with the study of the diffusion coefficient and the current fluctuations in a GEP where maximally two (interacting) particles can occupy each lattice site. The first three cumulants of the current distribution are calculated by combining the AP with the results from \cite{EPL2013}. In one and two dimensions the first three cumulants obtained from kMC are in agreement with the predicted values. In three dimensions the first two cumulants are in agreement with the AP. The statistics for the third cumulant is insufficient for a reliable comparison. Because the diffusion coefficient depends on the dimension, the current statistics change for different dimensions. The current statistics are independent of the spatial dimension for the SSEP and the zero-range process (ZRP).

The paper is organized as follows. In Section \ref{sec::theory} we introduce the quantities that are studied. It is explained how to predict the current distribution in any dimension from the AP. In Section \ref{sec::SSEP} we present the numerical results for the SSEP. The GEP is defined in Section \ref{sec::genmod}. The behavior of the diffusion coefficient in different dimensions is discussed in Section \ref{sec::diff}. Current fluctuations are studied in Section \ref{sec::fano}. A conclusion is presented in Section \ref{sec::conclude}.

\section{Theory\label{sec::theory}}

Consider a one-dimensional system of length $L$ in contact with two particle reservoirs, called $A$ and $B$, at densities $\rho_A$ and $\rho_B$. The dynamics in the bulk of the system is diffusive, i.e., there is no external driving in the bulk. The total number of particles that have passed through the system in the time interval $[0,t]$, in the NSS, is denoted by $Q_t$. To measure $Q_t$ one could, e.g., count the net number of particles entering the system from reservoir $A$. $Q_t$ is a stochastic quantity and is described by a probability distribution $P(Q_t)$. We study $P(Q_t)$ in the limit $t \uparrow \infty$ and $L \uparrow \infty$. Bodineau and Derrida showed that, by postulating an AP, one can calculate the cumulants of $P(Q_t)$ in a one-dimensional system from the integrals $I_m$ \cite{bodineau2004,bodineau2007cumulants}
\begin{equation}\label{eq::Ingen}
I_m = \int_{\rho_B}^{\rho_A} D(\rho) \sigma(\rho)^{m-1} d \rho.
\end{equation}
$D(\rho)$ is the diffusion coefficient. It is defined by Fick's first law
\begin{equation}\label{eq::fickfirst}
j  = - D(\rho) \frac{\Delta \rho}{L},
\end{equation}
where $j = \langle Q_t \rangle /t$ is the average particle flux ($\langle \cdot \rangle$ denotes the average over $P(Q_t)$), and with $\Delta \rho = \rho_B - \rho_A$ small enough so that linear response is valid. $\sigma(\rho)$ describes equilibrium fluctuations of $Q_t$ for large $t$ \begin{equation}
\frac{\langle Q_t^2 \rangle}{t} = \frac{1}{L} \sigma(\rho), \quad \rho_A = \rho_B = \rho.
\end{equation}
The first three cumulants of $P(Q_t)$ are equal to
\begin{eqnarray}
\frac{\langle Q_t \rangle}{t} = \frac{1}{L} I_1, \label{eq::I1} \\
\frac{\langle Q^2_t \rangle_c}{t} = \frac{\langle Q_t^2 \rangle - \langle Q_t \rangle^2}{t} = \frac{1}{L} \frac{I_2}{I_1}, \label{eq::I2} \\
\frac{\langle Q^3_t \rangle_c}{t} = \frac{\langle (Q_t - \langle Q_t \rangle)^3 \rangle}{t} = \frac{1}{L} \frac{3(I_3 I_1 - I_2^2)}{I_1^3}. \label{eq::I3}
\end{eqnarray}
The ratio of the first two cumulants is called the Fano factor
\begin{equation}\label{eq::fanoadd}
F = \lim_{L \rightarrow \infty} \lim_{t \rightarrow \infty} \frac{\langle Q_t^2 \rangle - \langle Q_t \rangle^2}{\langle Q_t \rangle} =  \frac{I_2}{I_1^2}.
\end{equation} 

Consider all density profiles $\rho_j(x,t')$, with $0 \leq x \leq L$ and $0 \leq t' \leq t$, that lead to the same particle flux $j$. In the long-time limit $t \uparrow \infty$, only the most probable (optimal) of these profiles is relevant for the current distribution \cite{bertini2014macroscopic}. The AP is valid as long as the optimal profiles are time-independent: $\rho_j(x,t') \equiv \rho_j(x)$. The point at which the optimal profile becomes time-dependent corresponds to a dynamical phase transition \cite{MFT_JSP_2006,PRE_bodineau_2005,PRL_hurtado_2011,PRE_espigares_2013}. For example, for one-dimensional systems on a ring, large fluxes are created by traveling waves \cite{PRE_bodineau_2005,PRL_hurtado_2011,PRE_espigares_2013}. One can show from the MFT that a sufficient condition on $D(\rho)$ and $\sigma(\rho)$ for the validity of the AP is \cite{MFT_JSP_2006}
\begin{equation}\label{eq::sufficient}
D(\rho) \sigma''(\rho) \leq D'(\rho) \sigma'(\rho), \quad \forall \rho.
\end{equation}
Note that \eref{eq::sufficient} is a sufficient but not a necessary condition.

A qualitative explanation of the AP goes as follows. The system is divided into subsystems. Their density profiles are considered to be independent of each other, except at the contacts between them. The subsystems should be so small that they are close to (local) equilibrium, but yet be large enough to allow for coarse graining. In this case, each subsystem has Gaussian current fluctuations around its deterministic behavior \eref{eq::fickfirst}, which is completely described by $D(\rho)$ and $\sigma(\rho)$. By calculating the optimal densities at the contacts between the subsystems, one finds the cumulant generating function (CGF) of the current distribution. From this CGF one can calculate \eref{eq::I1}, \eref{eq::I2}, \eref{eq::I3}. Hence, the AP allows one to calculate the current distribution arbitrarily far from equilibrium using only the equilibrium quantities $D(\rho)$ and $\sigma(\rho)$.

We now consider systems in $d \geq 1$ dimensions. Fick's first law is then given by
\begin{equation}
\vec{j} = - \mathbf{D}(\rho) \vec{\nabla} \rho,
\end{equation}
with $\mathbf{D}(\rho)$ a symmetric $d \times d$ matrix. 
If the diffusion is isotropic, which is the case considered here, one can write $\mathbf{D}(\rho) =  D_d(\rho) {\mathbb{I}}_d$, with $D_d(\rho)$ a scalar function depending on the dimension. A sufficient condition that excludes the possibility of a dynamical phase transition is \eref{eq::sufficient} with the scalar functions $D_d(\rho)$ and $\sigma_d(\rho)$ \cite{MFT_JSP_2006}. 

Akkermans et al.~studied current fluctuations in higher-dimensional diffusive systems \cite{EPL2013}. The shape of the system and the contacts with the reservoirs are taken arbitrary, but macroscopic in size. If the optimal density and current profiles are time-independent, the MFT predicts that the CGF of the system in $d$ dimensions $\mu_d(\lambda)$ equals \cite{EPL2013}
\begin{equation}\label{eq::mudmu1}
\mu_d(\lambda) = \kappa \mu_1(\lambda),
\end{equation}
with $\kappa$ a constant that depends on the shape of the system and shape of the contacts with the reservoirs. The calculation of $\kappa$ is explained in \ref{app::Cd}.
$\mu_1(\lambda)$ is the CGF of a one-dimensional system described by $D_d(\rho)$ and $\sigma_d(\rho)$. Since one assumes that the optimal density and current profiles are time-independent, $\mu_1(\lambda)$ can be calculated from the AP, by using $D_d(\rho)$ and $\sigma_d(\rho)$ in \eref{eq::Ingen}.
 
\section{Symmetric simple exclusion process\label{sec::SSEP}}

The SSEP is a stochastic lattice gas where particles interact by exclusion, i.e., each site can contain maximally one particle. Each particle attempts to hop to its nearest neighbors with unit rate. A hopping attempt is successful if the site is empty. The distance between two sites is equal to one. For the SSEP $D(\rho) = 1$ and $\sigma(\rho) = 2 \rho (1 - \rho)$ in any dimension. \eref{eq::sufficient} is therefore always satisfied. We consider reservoirs with densities $\rho_A = 1$ and $\rho_B = 0$. A calculation from the AP \cite{bodineau2004} or an exact microscopic derivation \cite{derrida2004current} shows that $F = 1/3$ in one dimension. Since $D(\rho)$ and $\sigma(\rho)$ are independent of the dimension, $F=1/3$ in any dimension. It is, however, important that the size of the contacts scales with the system size, thereby maintaining a finite fraction of the boundary in contact with the reservoirs. The numerical computation of the Fano factor in \cite{EPL2013} was performed for systems where this scaling is absent. We present simulations in which the contacts do scale with the system size.

\begin{figure}%
\centering
\subfloat[][]{
\includegraphics[width=0.3\columnwidth]{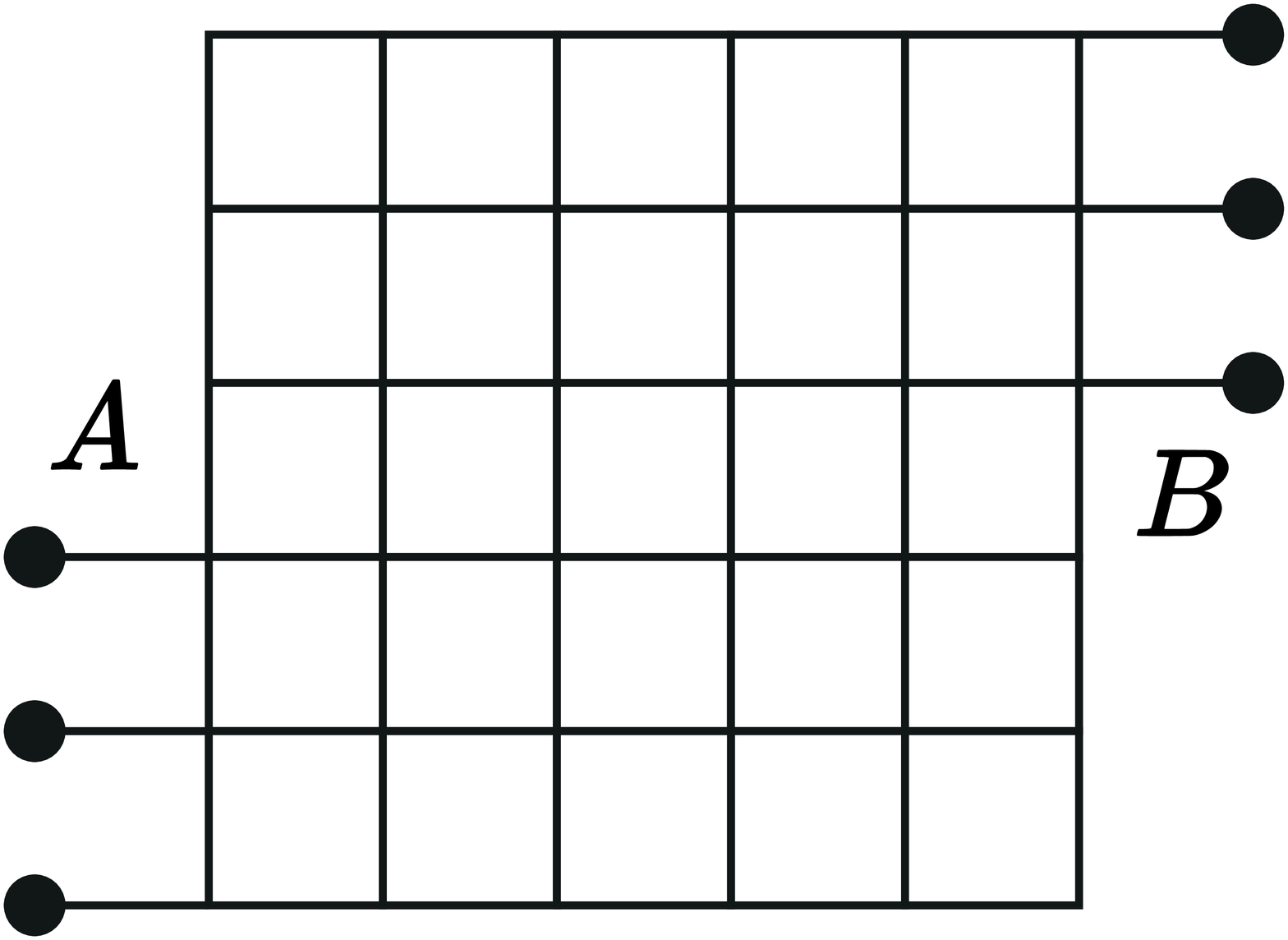}
\label{fig::SSEP_lattice_2D}
}%
\subfloat[][]{
\includegraphics[width=0.3\columnwidth]{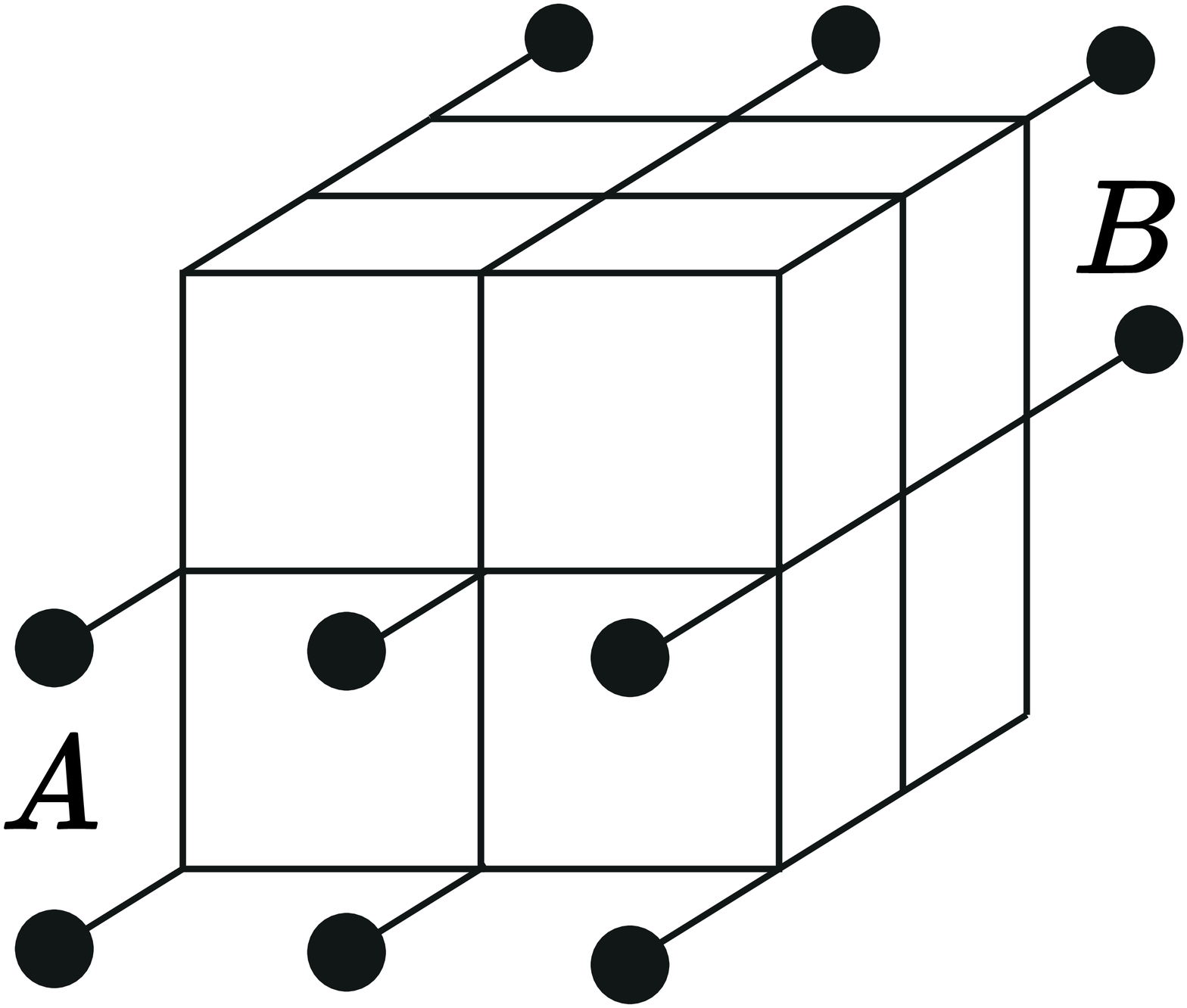}
\label{fig::SSEP_lattice_3D}
}
\caption{The type of contacts used for the SSEP in two dimensions $(a)$ and in three dimensions $(b)$. The black dots are sites with a particle density of 1 ($A$) or 0 ($B$), whose state is uncorrelated from the rest of the system. In two dimensions, $1/2$ of the lower left is connected to reservoir $A$ and $1/2$ of the upper right is connected to reservoir $B$. In three dimensions, $2/3$ of the lower left is connected to reservoir $A$ and $2/3$ of the upper right is connected to reservoir $B$.}%
\label{fig::SSEP_lattice}%
\end{figure}

The dynamics is simulated using a kMC algorithm, cf.~\ref{app::algorithm}. How the Fano factor is computed from the simulation data is explained in \ref{app::datacomp}. In two dimensions we consider squares of size $L \times L$ and in three dimensions cubes of size $L \times L \times L$. The contact between the system and the reservoirs is modeled as lattice sites whose densities are fixed and uncorrelated from the rest of the system, as in \cite{EPL2013}. The shape of the contacts is illustrated in Figure \ref{fig::SSEP_lattice}. 

\begin{figure}%
\subfloat[][]{
\includegraphics[width=0.47\columnwidth]{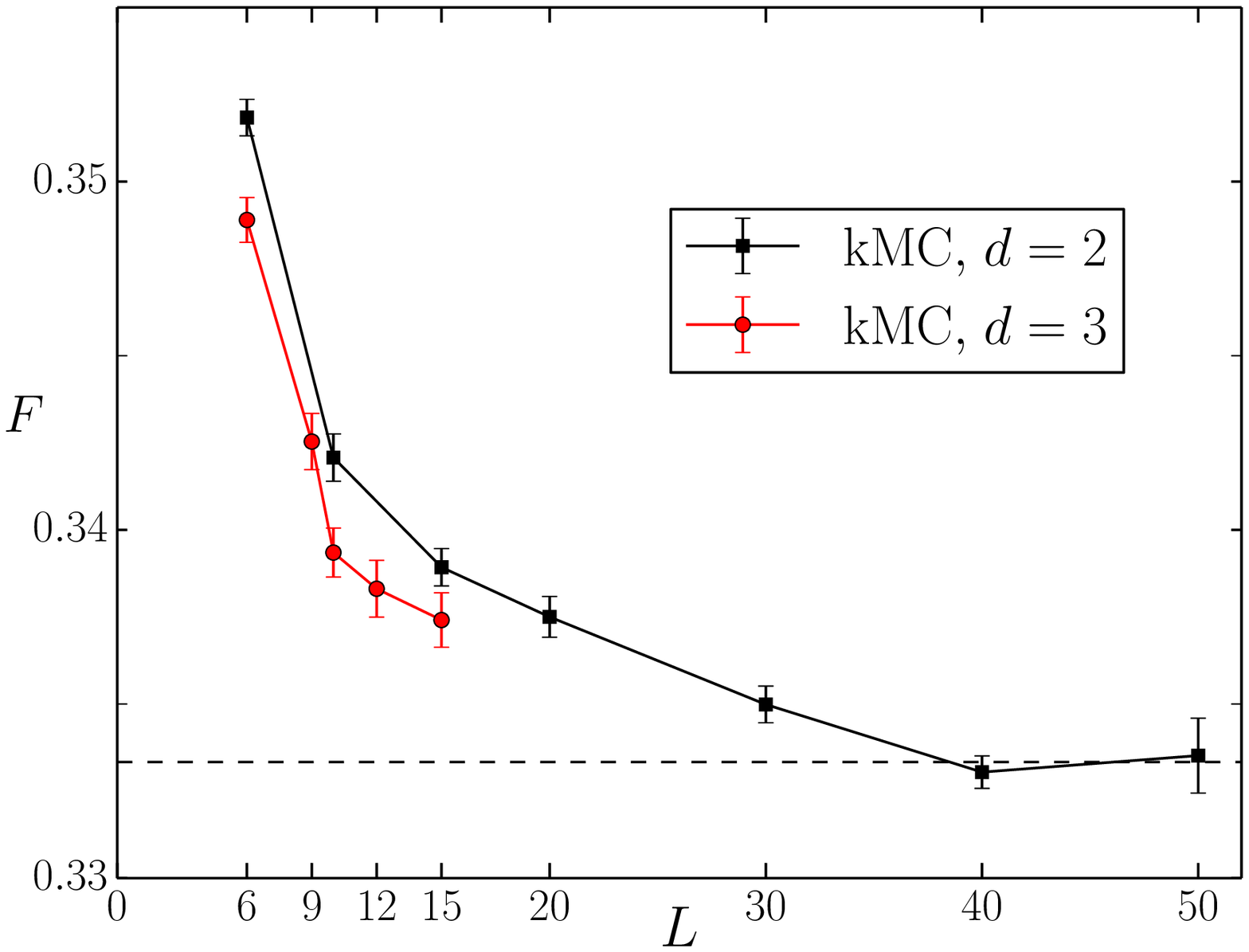}
\label{fig::fano_SSEP}
}%
\subfloat[][]{
\includegraphics[width=0.47\columnwidth]{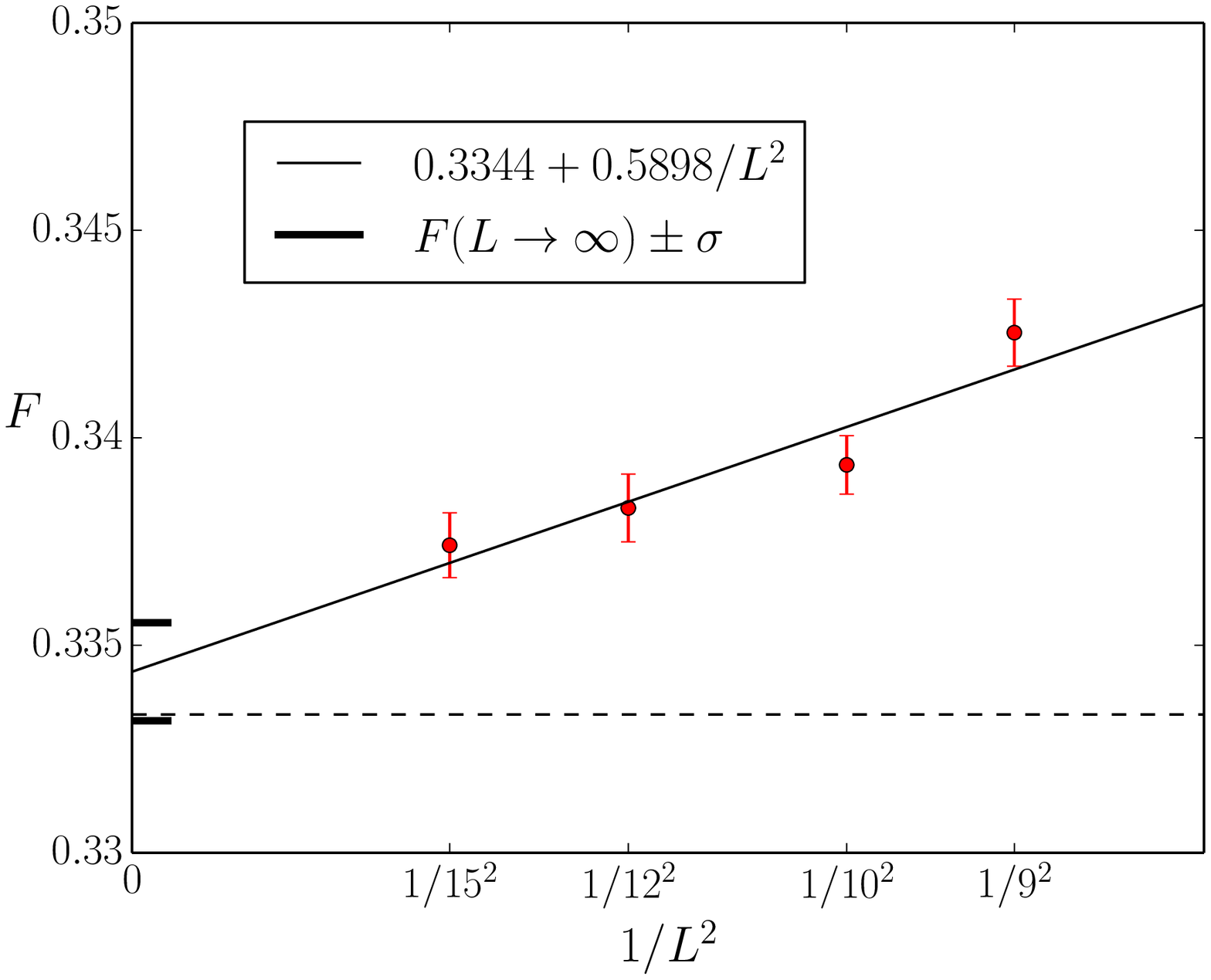}
\label{fig::fano_SSEP_3D}
}
\caption{(a) The Fano factor with one-sigma error bars for the SSEP, for squares $L \times L$ and cubes $L \times L \times L$ as depicted in Figure \ref{fig::SSEP_lattice}. The lines are a guide to the eye. The two-dimensional results show a convergence to $1/3$ at $L \approx 40$. The three-dimensional results have not yet converged. $(b)$ The three-dimensional data as a function of $1/ L^2$ for $L \geq 9$. The thin black line is a $1/L^2$ fit using the method of least squares with weighted error bars. The thick black lines are one-sigma error bars on the $L \rightarrow \infty$ limit predicted by the fit.}%
\label{fig:fano_ssep_all}%
\end{figure}

The numerical results for the Fano factor are presented in Figure \ref{fig::fano_SSEP}. For two dimensions the Fano factor has converged to 1/3 at $L \approx 40$. This extends the extrapolation presented in Figure 3 of \cite{EPL2013}. We determined numerically that $\kappa \approx 0.663 L$ for the geometry in Figure \ref{fig::fano_SSEP}, cf.~\ref{app::Cd}. The average current indeed converges to $L \langle Q_t \rangle / t \approx 0.663 L$, compared to $L \langle Q_t \rangle / t = 1$ in one dimension (data not shown).
For three dimensions convergence to 1/3 is not yet attained at $L = 15$. However, the data indicate convergence to 1/3 for larger system sizes. For the same distance $L$ between the two reservoirs, the Fano factor in three dimensions is lower than in two dimensions. One therefore expects convergence before $L = 40$ in three dimensions. In Figure \ref{fig::fano_SSEP_3D} we plot the Fano factor in three dimensions as a function of $1 / L^2$. There is no specific reason to assume that this is the correct convergence law. We choose this scaling because we want to compare our results with Figure 4 of \cite{EPL2013}. A $1/ L^2$ fit indicates an $L \rightarrow \infty$ limit of $F = 0.3344$, with one-sigma error bar $\sigma = 0.0018$. The fit was performed using the method of least squares with weighted error bars \cite{fit_book}. The extrapolation is in agreement with the expected value of $F=1/3$. Our numerical results validate the conjecture from \cite{EPL2013} that the observed discrepancy between numerics and theory is caused by too small contacts with the reservoirs.

\section{Generalized exclusion processes}

\subsection{The model \label{sec::genmod}}

\begin{figure}
\centering
\includegraphics[width=0.65\columnwidth]{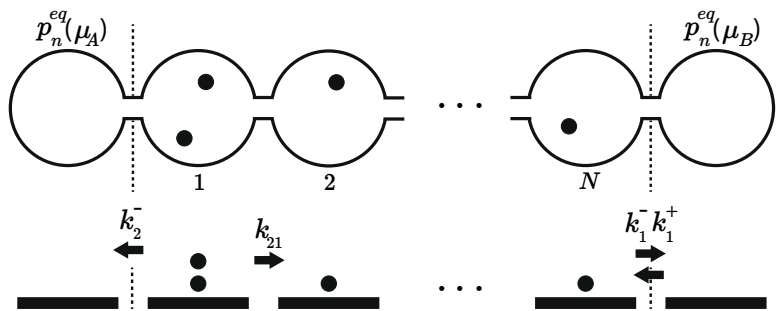}
\caption{A lattice model of a nanoporous material. Each cavity (upper drawing) is mapped to a lattice site (lower drawing) and contains between 0 and $\nmax$ particles (here $\nmax = 2$). On the boundaries the system is connected to cavities that are uncorrelated from the system. A cavity containing $n$ particles has equilibrium free energy $F(n)$.}
\label{fig::modelsys}
\end{figure}

Recently, we have studied the diffusive behavior of a lattice model of interacting particles \cite{PRL2013,PREbecker,EPJSTbecker}. The original motivation was the study of diffusion in nanoporous materials \cite{PRLchmelik2010}. Some of these materials have a structure consisting of cavities connected by narrow windows \cite{PCCPkrishna2013}, as illustrated in one dimension in Figure \ref{fig::modelsys}. Each cavity can be identified as a lattice site and can contain between 0 and $\nmax$ particles. The distance between two lattice sites is taken equal to one. The length of the system is the distance between the two reservoirs $L=N+1$, with $N$ the number of cavities. A cavity containing $n$ particles has an equilibrium free energy $F(n)$ that depends solely on the number of particles it contains. 
If the system is in equilibrium at chemical potential $\mu$ and inverse temperature $\beta = (k_b T)^{-1}$ (with $k_b$ the Boltzmann constant), the probability to observe $n$ particles in any cavity is equal to
\begin{equation}\label{eq::probeq}
p^{\mathrm{eq}}_n(\mu) = \left[ \mathcal{Z}(\mu)\right]^{-1} e^{- \beta \left[F(n)-\mu n\right]},
\end{equation}
with \( {\mathcal{Z}} \) the grand-canonical partition function
\begin{equation}
{\mathcal{Z}}(\mu) = \sum_{n=0}^{\nmax} e^{- \beta \left[F(n)-\mu n\right]}.
\end{equation}
Averages over the equilibrium distribution \eref{eq::probeq} are denoted by $\langle \cdot \rangle$, e.g.,
\begin{equation}\label{eq::defnav}
\nav (\mu) = \sum_{n=0}^{\nmax} n \peq.
\end{equation}
(Whether $\langle \cdot \rangle$ denotes the average over $\peq$ or $P(Q_t)$ is always clear from the context.) Particles jump from a cavity containing $n$ particles to a cavity containing $m$ particles with rate $k_{nm}$. These rates obey local detailed balance $p^{\mathrm{eq}}_n p^{\mathrm{eq}}_m k_{nm} = p^{\mathrm{eq}}_{m+1} p^{\mathrm{eq}}_{n-1} k_{m+1,n-1}$. 
The reservoirs are modeled as cavities whose probability distribution is uncorrelated from the rest of the system. 
The rates at which a reservoir cavity at chemical potential $\mu$ adds ($k^+_n$) or removes ($k^-_n$) one particle from a cavity containing $n$ particles are
\begin{equation}
k^+_n = \sum_{m=1}^{\nmax} k_{mn} p^{\mathrm{eq}}_m(\mu); \quad k^-_n = \sum_{m=0}^{\nmax-1} k_{nm} p^{\mathrm{eq}}_m(\mu).
\end{equation}

This model is a GEP \cite{kipnis} with a stochastic thermodynamical interpretation for the equilibrium statistics and dynamics. When defined like this it is an adequate model for the understanding of the equilibrium and diffusive behavior of particles in nanoporous materials \cite{PRL2013,PREbecker,EPJSTbecker}. For $\nmax = 1$ the model reduces to the SSEP. A zero-range process \cite{ZRPevansreview} is defined by rates that only depend on the cavity from which the particle jumps. Hence, one finds a ZRP for $\nmax = \infty$ and $k_{nm} = k_n$.

In the following, we fix the parameters to $\nmax = 2$ and $\beta = 1$. The free energy can be written as $F(n) = \ln n! + c n + f(n)$, with $c$ a constant \cite{PRL2013,PREbecker}. The first term accounts for the indistinguishability of the particles. The linear term $cn$ is the ideal gas contribution. $f(n)$ is nonzero because of particle interactions, and is called the interaction free energy. Note that a linear term in $F(n)$ is equivalent to adding a constant to the chemical potential $\mu$ \eref{eq::probeq}. A linear term does therefore not influence the equilibrium statistics at a given particle concentration. The rates we consider are
\begin{equation}\label{eq::ratesused}
k_{nm} = n e^{- \left[ f(n-1) + f(m+1) - f(n) - f(m) \right]/2}.
\end{equation}
It is clear that a linear term in $F(n)$ (or $f(n)$) also does not influence these rates. Hence, we can rescale $F(n)$ so that $f(0) = f(1) \equiv 0$ without loss of generality. All possible interactions are then described by $f(2)$. In the following we consider attractive particles $f(2) < 0$. In this case, the form of the transition rates \eref{eq::ratesused} can be rationalized from transition-state theory \cite{PREbecker}. Furthermore, for this choice of rates the diffusive behavior agrees with experiments of attractive particles in nanoporous materials \cite{PRL2013}.

For an isothermal system, which we consider here, $D(\rho)$ and $\sigma(\rho)$ are related by the following fluctuation-dissipation relation \cite{derrida2007non}
\begin{equation}\label{eq::sigmafdr}
\sigma(\rho) = 2 k_b T \rho^2 \kappa(\rho) D(\rho), 
\end{equation}
with $\kappa(\rho)$ the isothermal compressibility.
One knows from statistical physics that $\kappa(\rho)$ can be written as
\begin{equation}
\kappa(\rho) = \beta \frac{V}{\nav} \frac{\nvar}{\nav},
\end{equation}
with $V$ the volume in which the average $\nav$ and particle fluctuations $\nvar$ are measured. Because particles in different cavities do not interact, one can take the averages over one cavity, $V=1$ and $\rho = \nav$. One then finds for $\sigma (\rho)$ \eref{eq::sigmafdr}
\begin{equation}\label{eq::sigmafdr2}
\sigma(\rho) = 2  (\nvar) D(\rho).
\end{equation}
Regarding notation, since $\rho = \nav$ we use $\rho$ and $\nav$ interchangeably. Also, averages $\langle \cdot \rangle$ are a function of the chemical potential of the reservoirs. These can, however, be straightforwardly converted to densities via \eref{eq::defnav}. In this paper we write everything as a function of the density.

From \eref{eq::sigmafdr2} one finds that $I_m$ \eref{eq::Ingen} can be written as
\begin{equation}\label{eq::InAP}
I_m = \int_{\langle n \rangle_B}^{\langle n \rangle_A} D(\langle n \rangle)^m \left[ 2 (\nvar)  \right]^{m-1} d \langle n \rangle,
\end{equation}
where $\nav_{A}$ and $\nav_{B}$ are the average number of particles in, respectively, reservoir cavity $A$ and $B$.
One can compute $I_m$ by numerically simulating $D(\langle n \rangle)$ and analytically calculating $\langle n^2 \rangle - \langle n \rangle^2$ from $\peq$.

\subsection{Diffusion coefficient\label{sec::diff}}

We have studied $D(\rho)$ in this model both numerically and analytically \cite{PRL2013,PREbecker,EPJSTbecker}. From these studies one can conclude that $D(\rho)$ is, in general, influenced by correlations (see also \cite{comment_becker}). Since the effect of correlations changes and is actually expected to diminish with increasing dimension, the function $D(\rho)$ depends on the dimension \cite{PREbecker,comment_becker}. If the effect of correlations upon the diffusion is completely neglected one can show that $D(\rho)$ is given by \cite{PRL2013,PREbecker}
\begin{equation}\label{eq::DtDMF}
D(\rho) = \frac{\langle k \rangle}{\nvar}, \qquad \quad \langle k \rangle = \sum_n \sum_m p^{\mathrm{eq}}_n p^{\mathrm{eq}}_m k_{n m}.
\end{equation}
This result is valid for a (hyper)cubic lattice in any dimension. Because one arrives at \eref{eq::DtDMF} by neglecting all correlations, it could be argued that in the limit of infinite dimension $D(\rho)$ converges to \eref{eq::DtDMF}. Although we do not have a rigorous proof of this statement, it is confirmed by numerical evidence given below (see also \cite{PREbecker}). We therefore denote the results that are calculated from \eref{eq::DtDMF} as the $d \rightarrow \infty$ limit. Note that in this limit the integral \eref{eq::InAP} can be calculated analytically.

The uncorrelated result \eref{eq::DtDMF} is exact for the SSEP ($\nmax = 1$), which is easily checked by using that $p^{\mathrm{eq}}_1 = \rho$ and $p^{\mathrm{eq}}_0 = 1 - \rho$. It is also the same in any dimension \cite{kutner1981chemical}. \eref{eq::DtDMF} is also exact for the one-dimensional ZRP \cite{EPJSTbecker}. Since the particle distribution in the NSS factorizes in any dimension for the ZRP \cite{evans2006}, the calculation from \cite{EPJSTbecker} can be straightforwardly extended to higher dimensions to show that $D(\rho)$ is independent of the dimension. To our knowledge, these are the only two cases where the uncorrelated result is exact for GEPs. It is, then, no surprise that $D(\rho)$ is independent of the dimension.

\begin{figure}
\centering
\includegraphics[width=0.52\columnwidth]{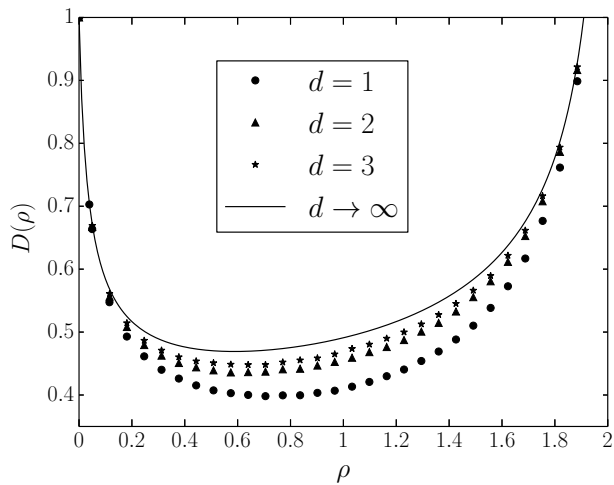}
\caption{$D(\rho)$ for $f(2) = - 2.5$ and $\nmax = 2$ in one, two, three, and infinite dimensions. The error bars are smaller than the symbol sizes.}
\label{fig::dt_cluster}
\end{figure}

\begin{figure}
\centering
\includegraphics[width=0.52\columnwidth]{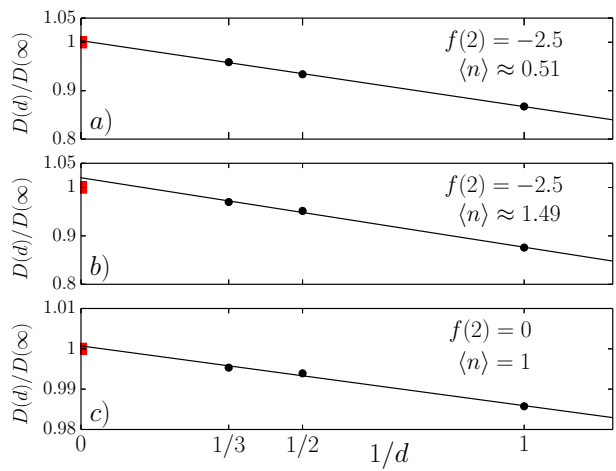}
\caption{The diffusion coefficient as a function of the dimension, for $\nmax = 2$. The data are normalized w.r.t.~the analytical uncorrelated result \eref{eq::DtDMF}, which is denoted by $D(\infty)$. The black circles are from kMC simulations and the red squares are \eref{eq::DtDMF}. The error bars are smaller than the symbol sizes. $1/d$ fits were performed with the method of least squares. In all three cases this fit provides a good estimate for the diffusion coefficient at infinite dimension, with a relative error ($D_{\mathrm{fit}}(\infty)/D(\infty) -1$) of $a)$ 0.3 \%, $b)$ 2.0 \%, and $c)$ 0.07 \%.}
\label{fig::Dt_1overd}
\end{figure}

We consider now $f(2) = - 2.5$. This is a concave $f(n)$, signifying attractive particles \cite{PRL2013}. We choose this interaction because correlations have a large effect for attractive particles. In Figure \ref{fig::dt_cluster} we plot $D(\rho)$ in one, two, three, and infinite dimensions. We refer to \ref{app::simD} for details on the simulations. $D(\rho)$ appears to converge with increasing dimension towards the $d \rightarrow \infty$ result \eref{eq::DtDMF}. The diffusion coefficient as a function of the dimension for $\nav \approx 0.51$ and $\nav \approx 1.49$ is shown in, respectively, Figures \ref{fig::Dt_1overd}a and \ref{fig::Dt_1overd}b. The behavior is well approximated by a $1/d$ dependence. Figure \ref{fig::Dt_1overd}c shows the same quantity for the interaction $f(2) = 0$ at $\nav  = 1$ (data from \cite{comment_becker}). Also here an approximate $1/d$ dependence is found. This dependence can be understood as follows. Correlations are the result of memory effects in the environment \cite{PREbecker}. The strongest contribution comes from the increased probability that a particle jumps back to its previous position. The probability to do so is approximately $1/2d$ as there are $2d$ neighboring cavities. This simple argument indeed suggests that the effect of correlations will decrease approximately as $\propto 1/d$. 

\subsection{Current fluctuations\label{sec::fano}}

\begin{figure}
\centering
\includegraphics[width=0.6\columnwidth]{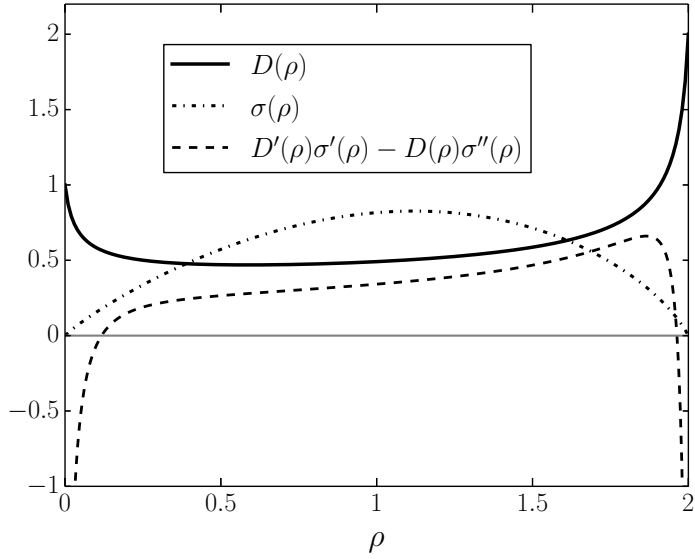}
\caption{Plot of $D(\rho)$, $\sigma(\rho)$, and $D'(\rho) \sigma'(\rho) - D(\rho) \sigma''(\rho)$ for $f(2) = - 2.5$ and $\nmax = 2$ in the limit $d \rightarrow \infty$ \eref{eq::DtDMF}. The sufficiency condition \eref{eq::sufficient} is satisfied if $D'(\rho) \sigma'(\rho) - D(\rho) \sigma''(\rho) \geq 0$ for all $\rho$.}
\label{fig::suff_cond}
\end{figure}

The sufficiency condition \eref{eq::sufficient} is not satisfied for $f(2) = -2.5$, as shown in Figure \ref{fig::suff_cond} for $d \rightarrow \infty$ \eref{eq::DtDMF}. The numerically simulated $D(\rho)$'s do not give smooth results for \eref{eq::sufficient}, since one has to calculate the second derivative of an interpolated function. The qualitative behavior of \eref{eq::sufficient} for finite dimensions is, however, the same as for $d \rightarrow \infty$. Starting from \eref{eq::DtDMF}, one sees that \eref{eq::sufficient} does not hold for many GEPs. One can show analytically that all GEPs with $\nmax = 2$ and $f(2) < 0$ do not satisfy \eref{eq::sufficient}. Numerically, one finds that GEPs with $\nmax=2$ and $f(2) \gtrsim 2.917$ also do not satisfy \eref{eq::sufficient}.
Although \eref{eq::sufficient} is not satisfied for the parameters considered here, we expect that the AP is still valid. Dynamical phase transitions have only been observed for closed systems \cite{MFT_JSP_2006,PRE_bodineau_2005,PRL_hurtado_2011,PRE_espigares_2013}, not boundary driven ones \cite{PRLhurtado,PRE_Hurtado_AP,PRE_mieke_2012}. Also, dynamical phase transitions do not occur for currents close to the average current \cite{bertini2014macroscopic}. Currents created by time-dependent density profiles, if any, are therefore highly unlikely, and their influence on the first three moments of the current distribution is expected to be negligible.

\begin{figure}%
\subfloat[][]{
\includegraphics[width=0.5\columnwidth]{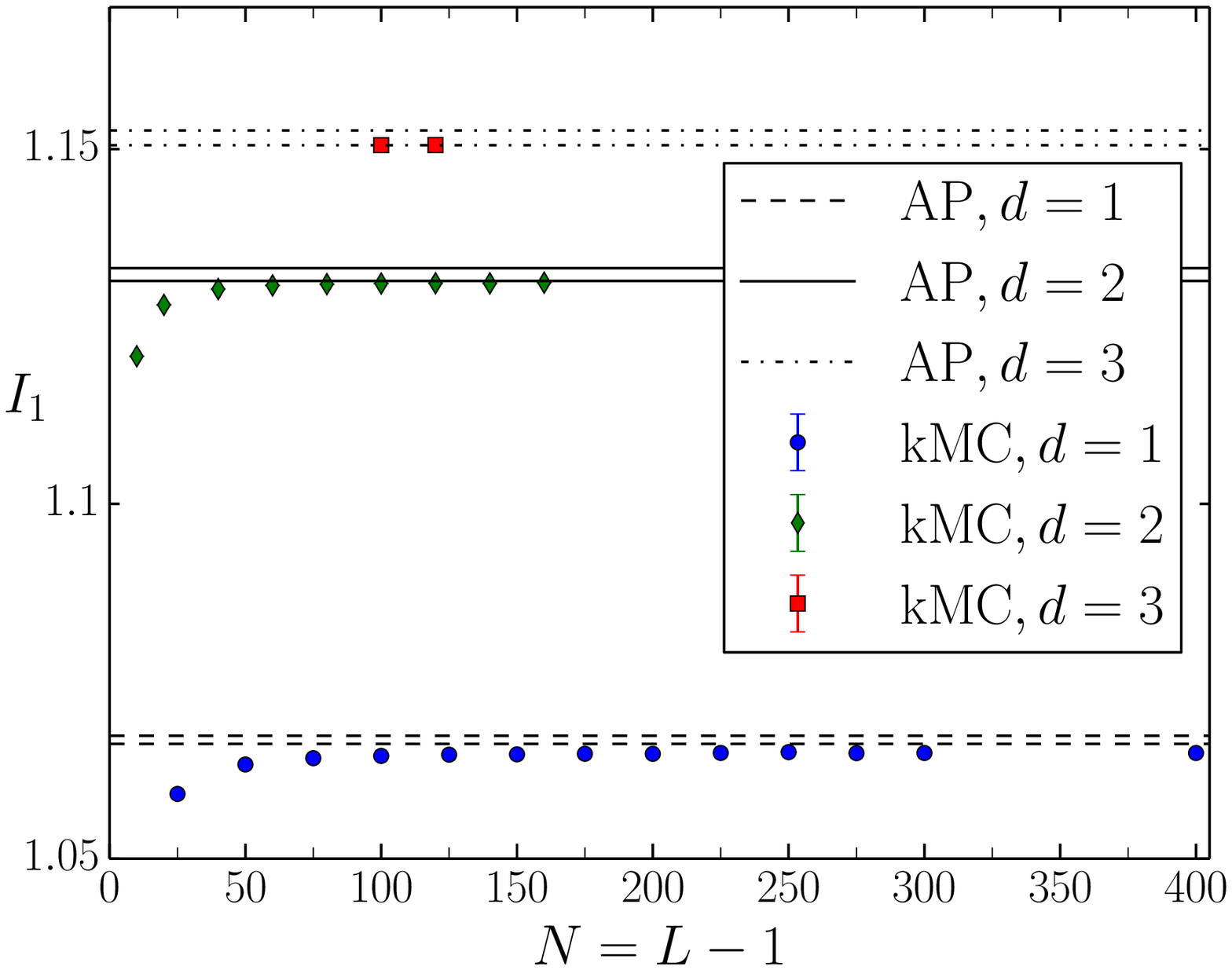}
\label{fig::I1_cluster}
}%
\subfloat[][]{
\includegraphics[width=0.5\columnwidth]{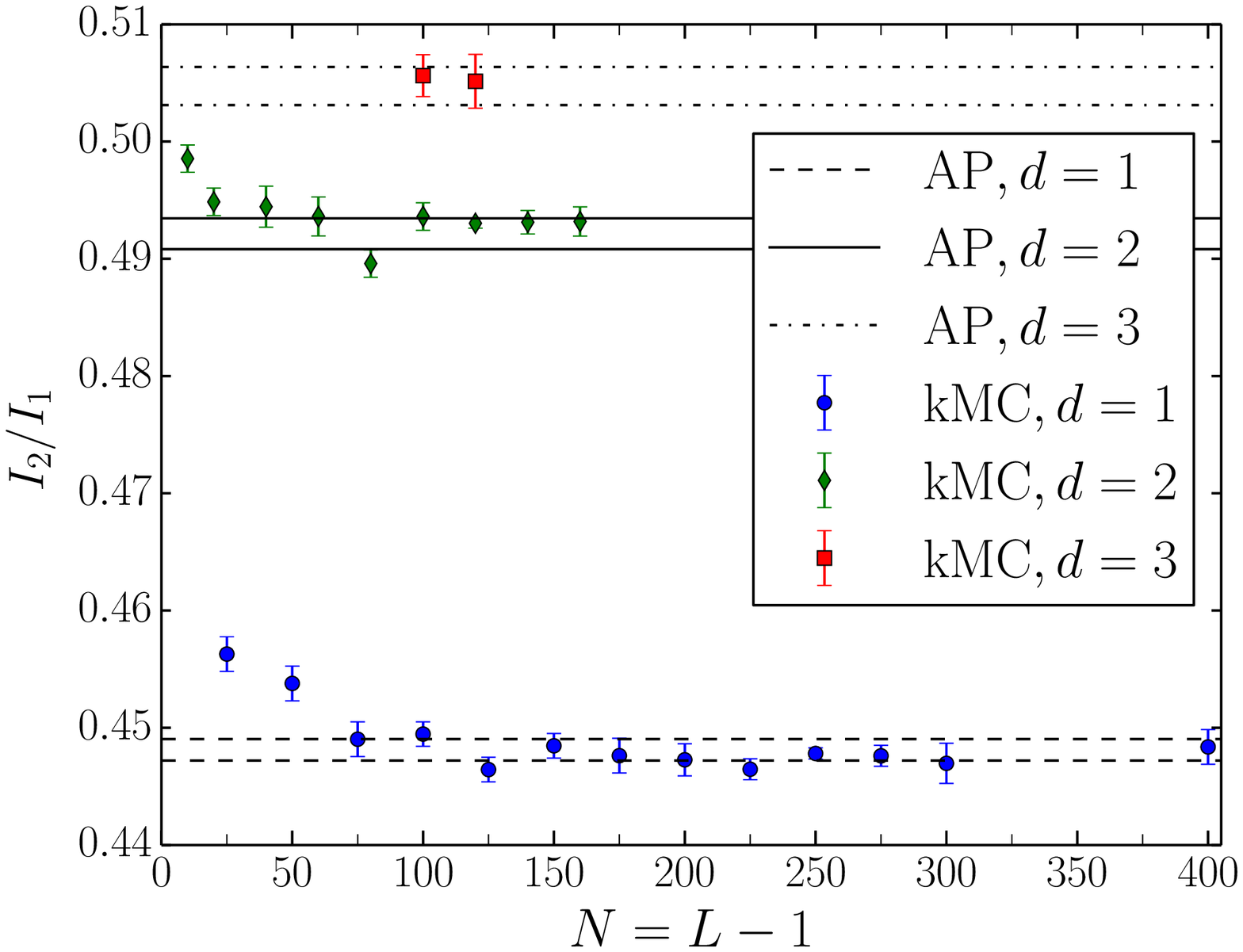}
\label{fig::var_cluster}
}
\caption{(a) $L \langle Q_t \rangle / t = I_1$ and (b) $L \langle Q^2_t \rangle_c / t = I_2 / I_1$, for $f(2) = -2.5$, $\nmax = 2$, and different lengths in one, two, and three dimensions. The lines are predictions from the AP, and represent one-sigma error bars. The points with one-sigma error bars are from a direct simulation of the current. In two (three) dimensions, the directly measured cumulants are divided by $L_y$ ($L_y L_z$), see \ref{app::Cd}. 
}
\label{fig::j_all}%
\end{figure}

We study the current statistics for $f(2) = - 2.5$ and  reservoir densities $\nav_A = \nmax = 2$ ($\mu_A = \infty$) and $\nav_B = 0$ ($\mu_B = - \infty$). Plots of $L \langle Q_t \rangle/t = I_1$ and  $L \langle Q^2_t \rangle_c /t = I_2/I_1$ as a function of the length are shown in, respectively, Figures \ref{fig::I1_cluster} and \ref{fig::var_cluster}. The values predicted by the AP are given by lines, which are the one-sigma error bars. These error bars arise from the error bars on the simulated $D(\rho)$'s. Values from direct numerical simulations are also given with one-sigma error bars, as explained in \ref{app::datacomp}. 

\begin{figure}%
\subfloat[][]{
\includegraphics[width=0.5\columnwidth]{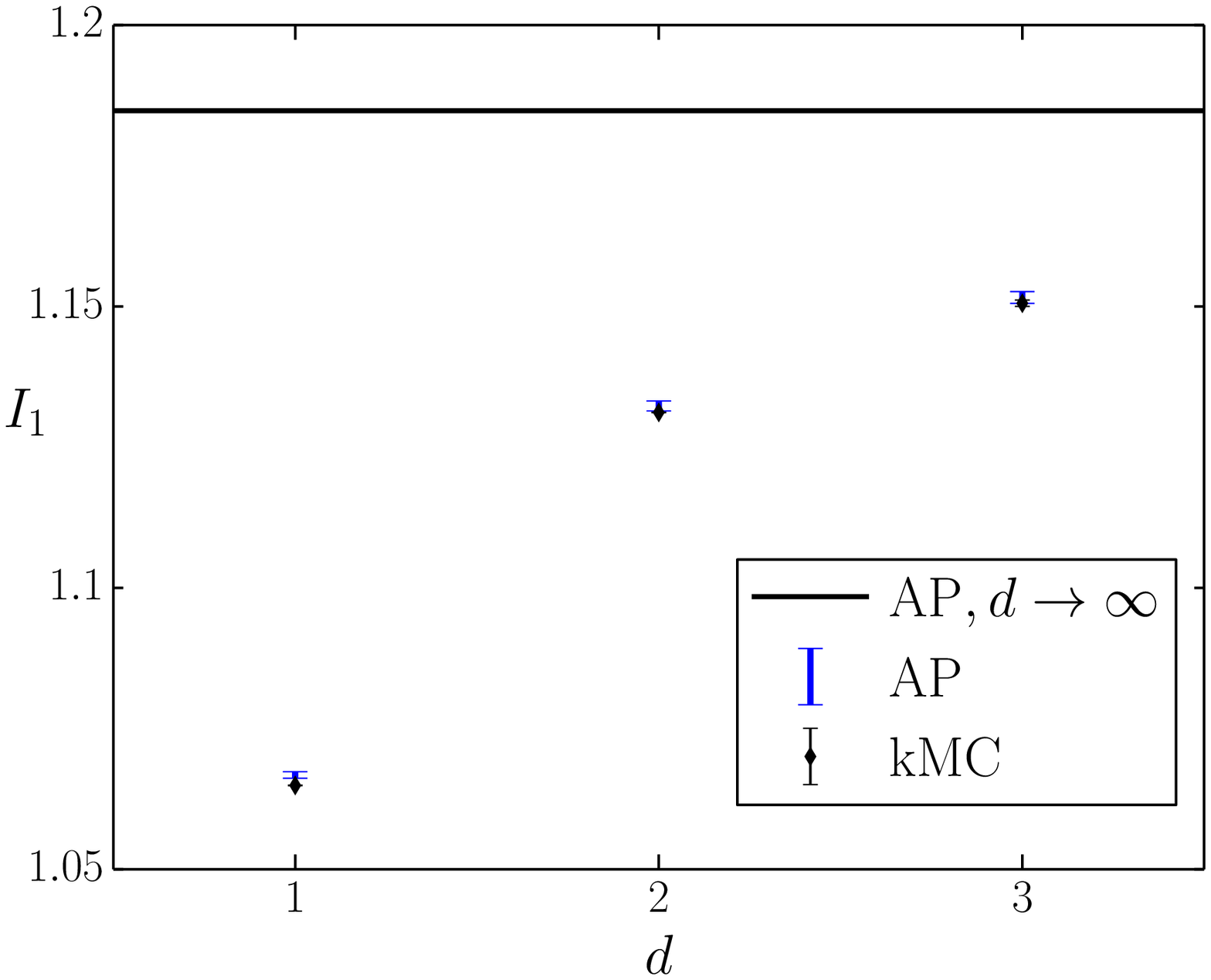}
\label{fig::I1_vs_dim}
}
\subfloat[][]{
\includegraphics[width=0.5\columnwidth]{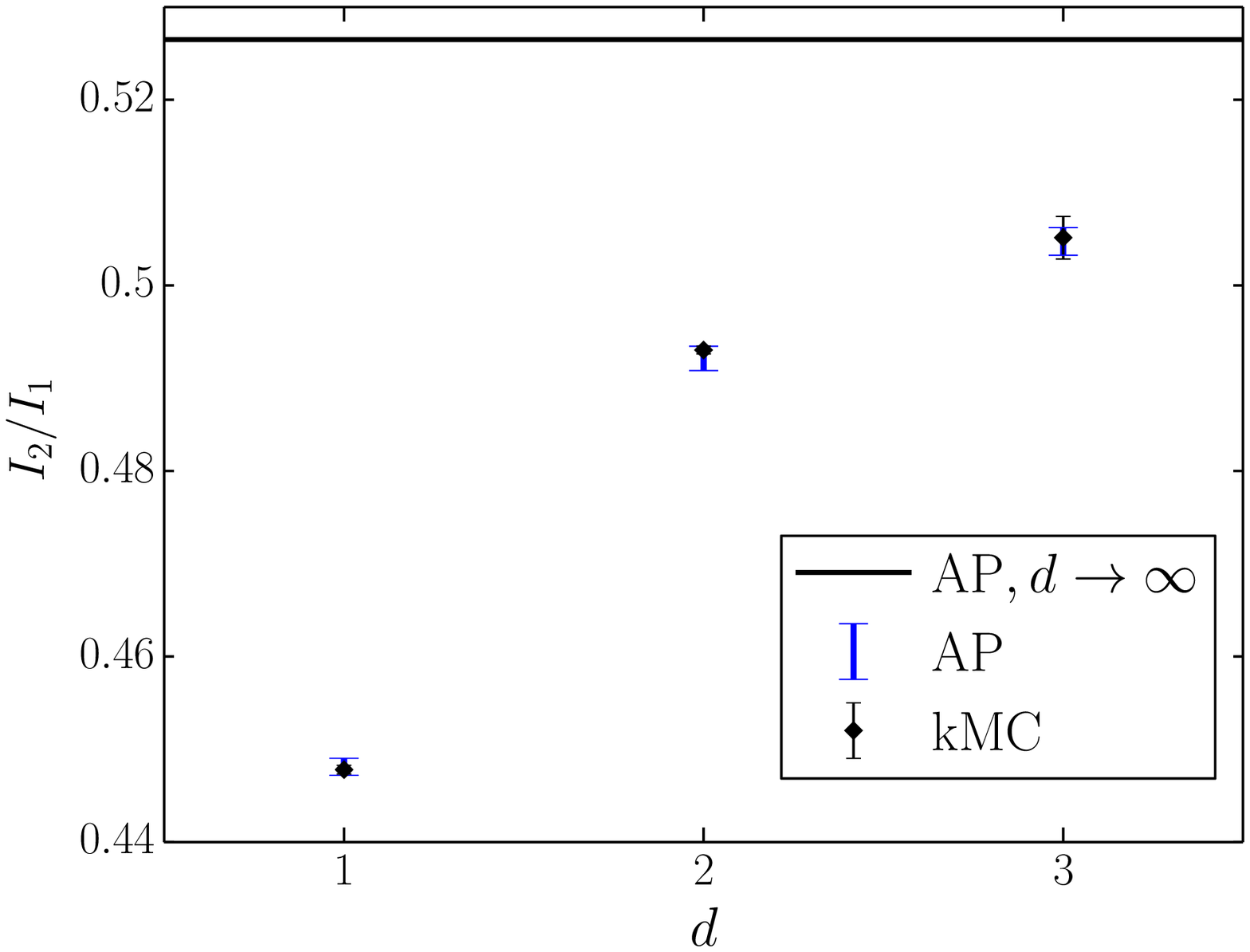}
\label{fig::Var_vs_dim}
} 
\\
\subfloat[][]{
\includegraphics[width=0.5\columnwidth]{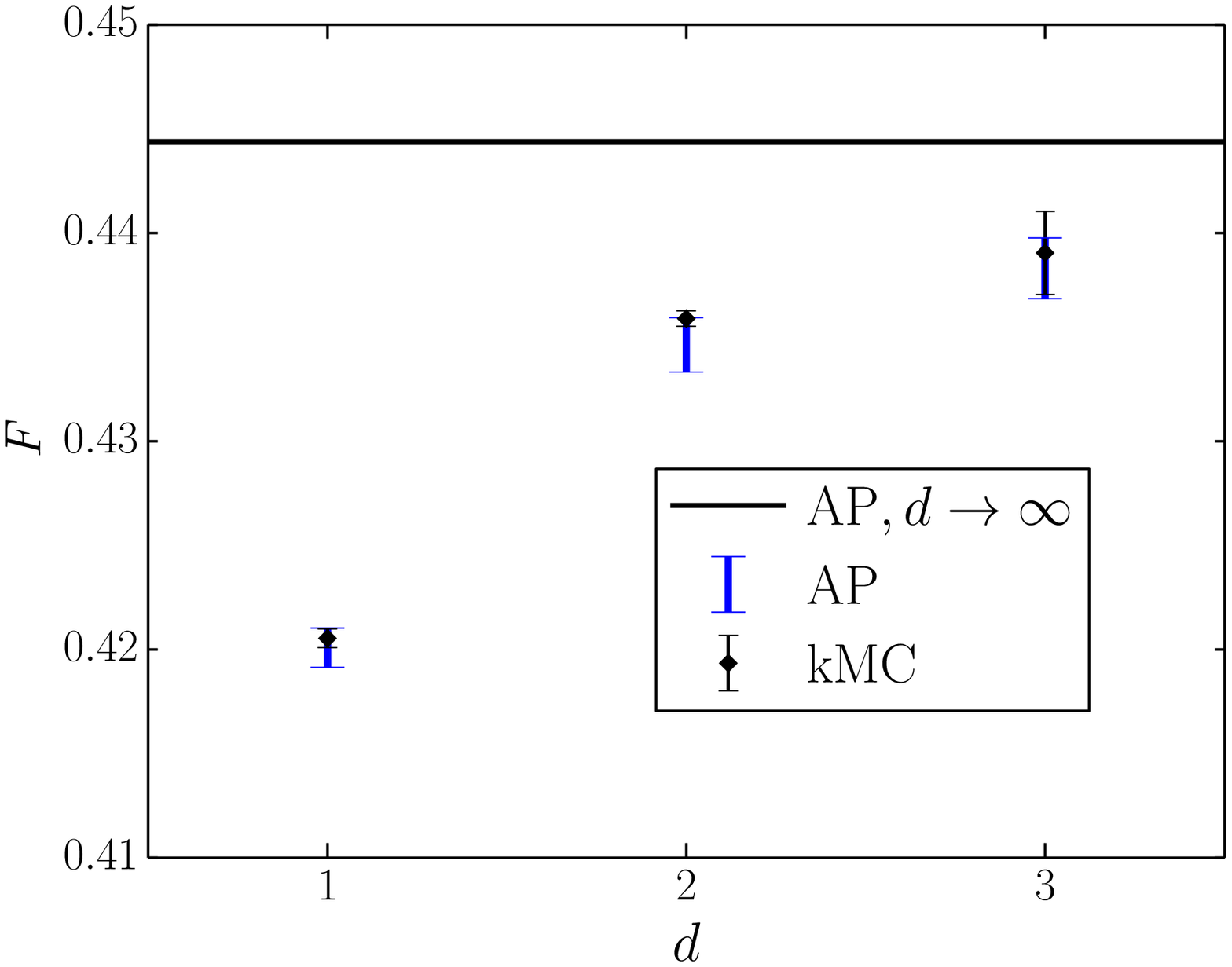}
\label{fig::Fano_vs_dim}
}
\subfloat[][]{
\includegraphics[width=0.5\columnwidth]{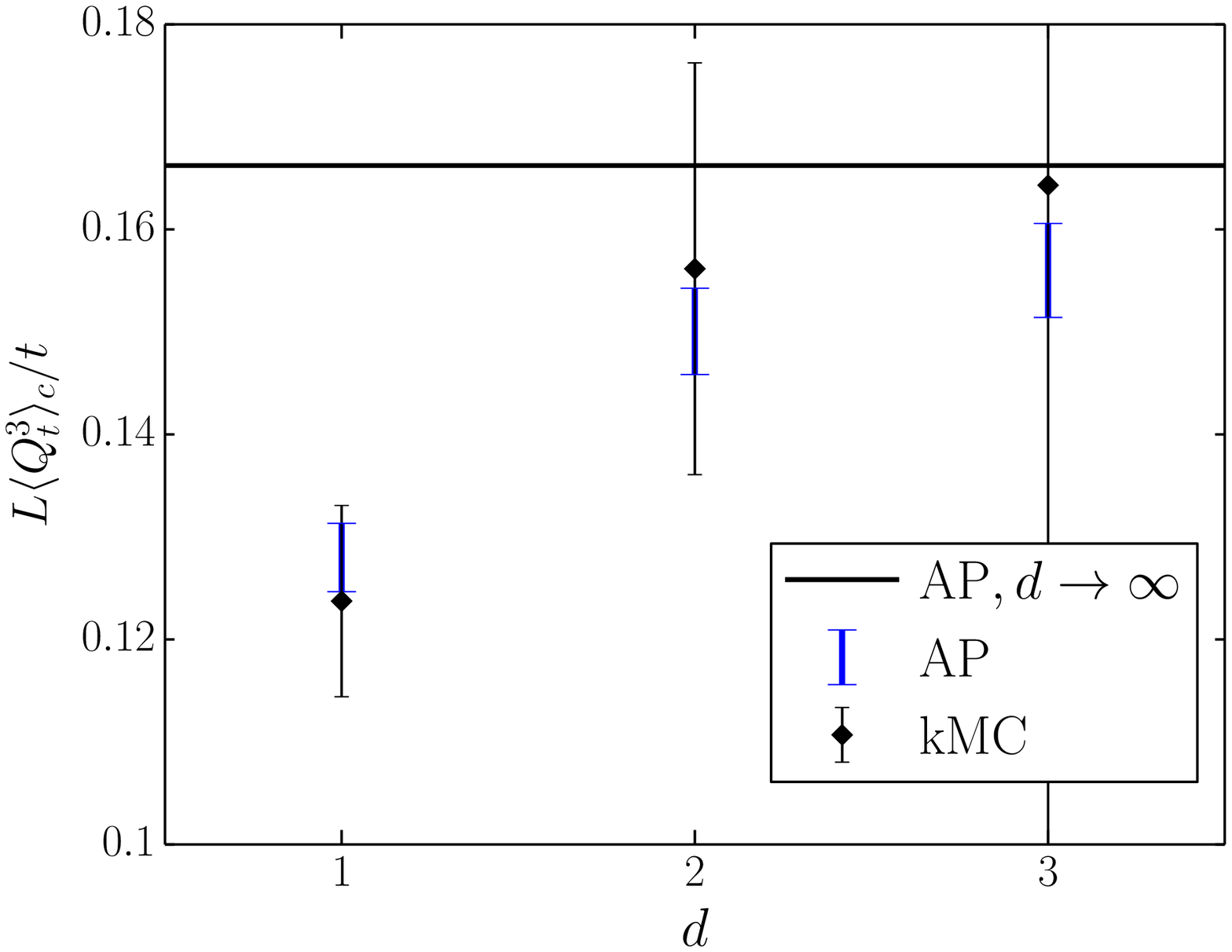}
\label{fig::C3_vs_dim}
}
\caption{(a) $I_1$, (b) $I_2/I_1$, (c) $F$, (d) $L \langle Q^3_t \rangle_c / t$ for $f(2) = -2.5$ and $\nmax = 2$ as a function of the dimension. Predictions from the AP are denoted by blue error bars without symbol. The limiting case $d \rightarrow \infty$ is shown as a black line. Direct numerical simulations are denoted by black diamonds. In two (three) dimensions, the directly measured cumulants are divided by $L_y$ ($L_y L_z$), see \ref{app::Cd}. Note that the error bar at $d=3$ in (d) spans approximately 7 times the whole $y$ axis.
}
\label{fig::dim_all}
\end{figure}

\begin{figure}%
\subfloat[][]{
\includegraphics[width=0.5\columnwidth]{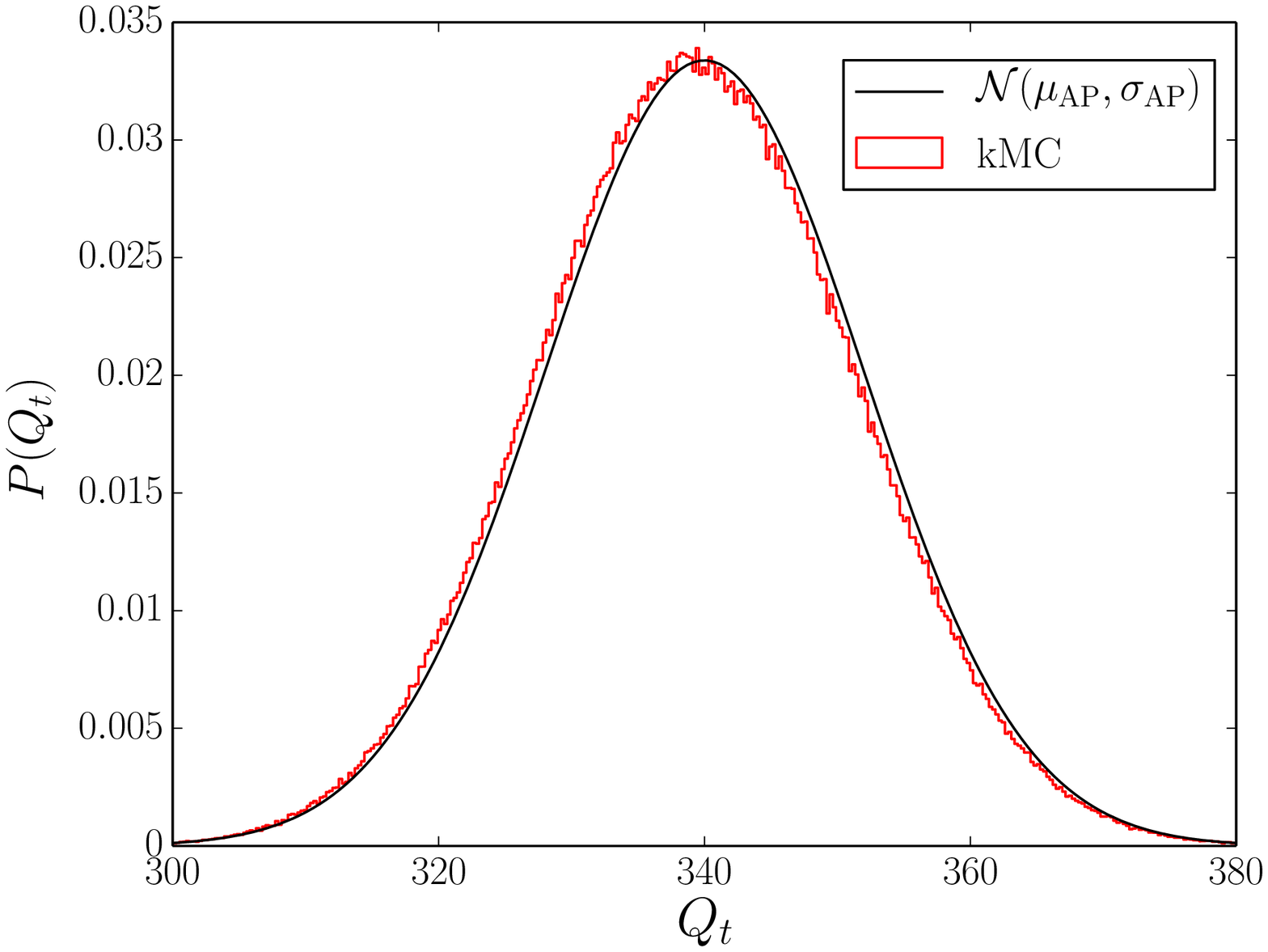}
\label{fig::histo_AP}
}%
\subfloat[][]{
\includegraphics[width=0.5\columnwidth]{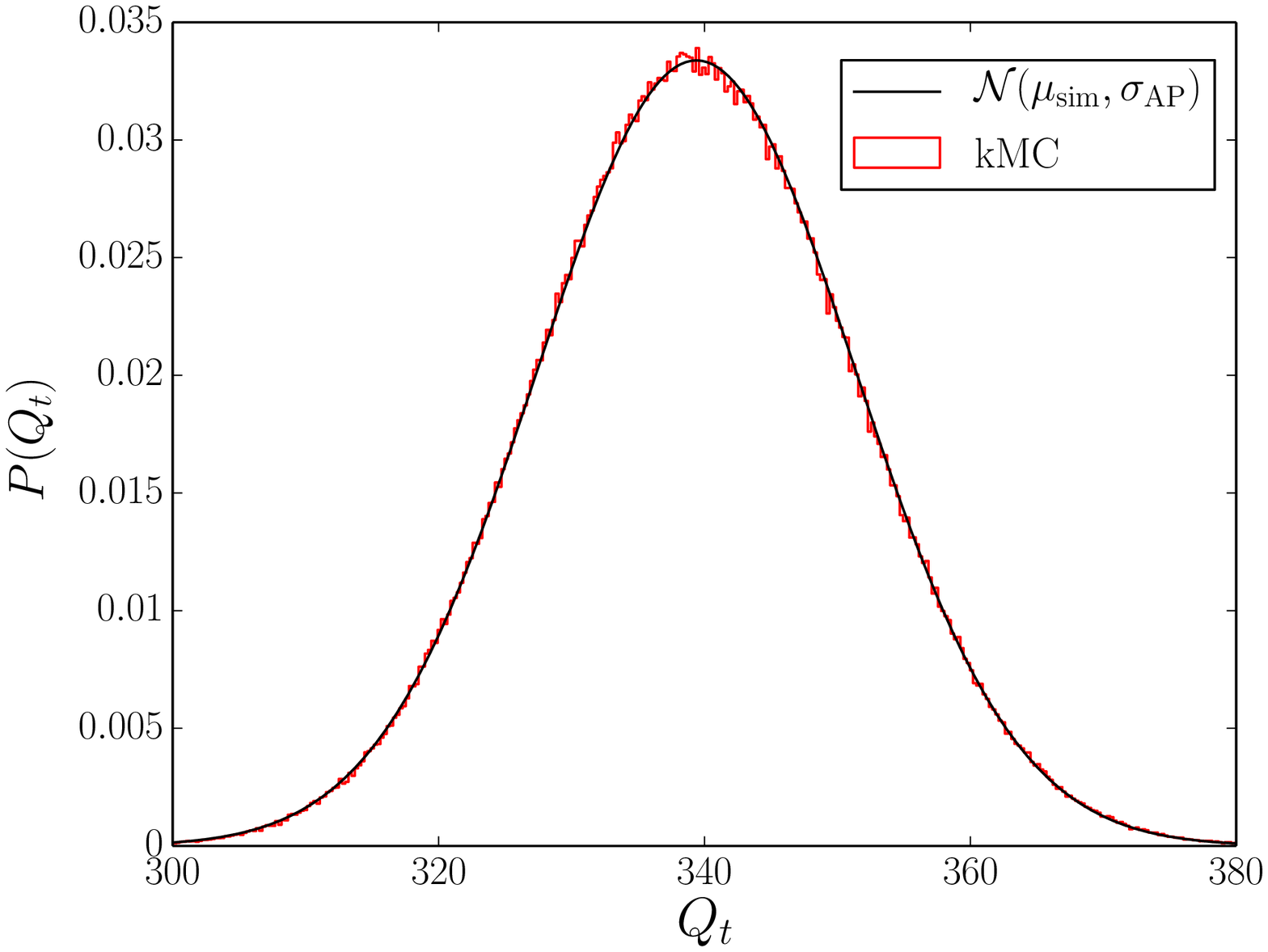}
\label{fig::histo_sim}
}
\caption{
$P(Q_t)$ from kMC (red) in one dimension for $\nmax = 2$, $f(2) = -2.5$, $L=251$, and $t=8.10^4$. A Gaussian distribution (black) with average and variance predicted by the AP (a) and the simulated average and variance from the AP (b) is also plotted. The data is well approximated by a Gaussian distribution.
}
\label{fig::histo}
\end{figure}

Let us first consider the one-dimensional data. We estimate convergence in length at $L \approx 175$. How we check for convergence in time is explained in \ref{app::datacomp}. The value for $I_1$, cf.~Figure \ref{fig::I1_vs_dim}, is taken from the highest considered length in Figure \ref{fig::I1_cluster}. To achieve a good statistics for the second and third cumulant, we have performed an extensive simulation for length $L = 251$. The simulated values for $L \langle Q^2_t \rangle_c /t = I_2/I_1$, $F = I_2/ I^2_1$, and $L \langle Q^3_t \rangle_c /t$ for this length are given in, respectively, Figures \ref{fig::Var_vs_dim}, \ref{fig::Fano_vs_dim}, and \ref{fig::C3_vs_dim}.
$I_1$ from the AP is slightly higher than the directly simulated value ($I^{\mathrm{AP}}_1 / I^{\mathrm{sim}}_1 \approx 1.0018$). The most likely reason for this is that the simulated $D(\rho)$ slightly overestimates the real $D(\rho)$. The diffusion coefficient should be measured in the limit of an infinitely small concentration gradient, while of course the simulations are performed at a finite concentration gradient. Similarly, one should in principle simulate an infinitely long system, so that all boundary effects have disappeared. Both approximations cause the numerically simulated $D(\rho)$ to overestimate the real value \cite{comment_becker}. Furthermore, to calculate $I_1^{\mathrm{AP}}$ one has to interpolate the simulated points of $D(\rho)$, and then integrate this interpolated function. This could introduce a small numerical imprecision. Since the relative difference is less than $0.2 \%$ we consider this result a very good agreement between $I^{\mathrm{AP}}_1$ and  $I^{\mathrm{sim}}_1$. Also the variance and the Fano factor are in very good agreement with the value from the AP: $(I_2^{\mathrm{AP}}/I_1^{\mathrm{AP}})/(I_2^{\mathrm{sim}}/I_1^{\mathrm{sim}}) \approx 1.0007$ and $F^{\mathrm{AP}} / F^{\mathrm{sim}} \approx 0.9989$. 

Figure \ref{fig::C3_vs_dim} shows the third cumulant. Although the error bars are significantly larger compared to the first two cumulants, the data indicate agreement between the AP and the directly simulated values. 
Finally, we plot $P(Q_t)$ obtained from kMC together with the Gaussian prediction from the first two moments of the AP in Figure \ref{fig::histo}. The small error on $I_1$ from the AP is noticeable for determining $\langle Q_t \rangle$. When using the simulated $\langle Q_t \rangle$, one sees that $P(Q_t)$ is well approximated by a Gaussian. Indeed, the skewness of $P(Q_t)$ is small $\langle Q^3_t \rangle_c / \langle Q^2_t \rangle^{3/2}_c \approx 0.034$, i.e., $P(Q_t)$ is almost symmetric. Although the difference is small, one observes that for $Q_t < 320$ the simulated $P(Q_t)$ is consistently lower than the Gaussian, while for $Q_t > 365$ it is consistently higher.

We now discuss the higher-dimensional systems. In contrast to the SSEP, all sites at the boundaries are in contact with the reservoirs. If periodic boundary conditions are imposed in the $y$ direction, $D(\rho)$ converges in two dimensions to the $L_y \uparrow \infty$ limit at $L_y \approx 3$. In the simulations we take $L_y = L_z = 5$ with periodic boundary conditions. The diffusion coefficient is simulated for the same concentration gradients and length in the $x$ direction as for the one-dimensional case. The different coupling to the reservoirs compared to the SSEP is done for numerical reasons. The program for the GEP is too slow to simulate a convergence in both the $x$ direction and $y$ ($z$) direction. The periodic boundary conditions employed here are equivalent to the $L_y (L_z) \uparrow \infty$ limit. All sites at the boundaries are coupled to the reservoirs because this gives the highest particle flux. The higher the particle flux, the better the current statistics for a given simulation time. If all boundary sites are connected to the reservoirs $\kappa = L_y$ and $\kappa = L_y L_z$ in, respectively, two and three dimensions. This is explained in \ref{app::Cd}.

For two dimensions we assume convergence in length at $L \approx 120$. The error on $I_1$ is comparable to the one-dimensional case ($I^{\mathrm{AP}}_1 / I^{\mathrm{sim}}_1 \approx 1.0010$). The second and third cumulants are determined from extensive simulations at length $L = 121$. The variance and Fano factor are slightly underestimated by the AP: $(I_2^{\mathrm{AP}}/I_1^{\mathrm{AP}})/(I_2^{\mathrm{sim}}/I_1^{\mathrm{sim}}) \approx 0.9982$ and $F^{\mathrm{AP}} / F^{\mathrm{sim}} \approx 0.9971$. We consider this a very good agreement between the direct simulations and predictions from the AP. The relative difference is less than 0.3 \%, and all quantities show a large overlap within their error bars. The third cumulant is also compatible with the AP prediction, although the error bar on the directly simulated value is rather large. The shape of $P(Q_t)$ is similar to the one-dimensional case (data not shown).

For three dimensions the simulation times become much longer. We therefore only simulate the current for systems of length $L = 101$ and $L=121$. Since the two-dimensional system has converged at $L=121$, one can safely assume that this is also the case for the three-dimensional system. The cumulants from Figure \ref{fig::dim_all} are calculated for $L=121$. The average, variance, and Fano factor are correctly predicted by the AP. There is insufficient data to achieve a reliable estimate for the third cumulant. The obtained value shown in Figure \ref{fig::C3_vs_dim} agrees well with the AP, but the error bar is very large: $L \langle Q^3_t \rangle_c /t + \sigma = 0.438$ and $L \langle Q^3_t \rangle_c /t - \sigma = -0.110$. The shape of $P(Q_t)$ is similar to the one-dimensional case (data not shown).

\section{Conclusion\label{sec::conclude}}

To conclude, we have studied numerically current fluctuations in the symmetric simple exclusion process (SSEP) and a generalized exclusion process (GEP). For the SSEP we find that the Fano factor is independent of the spatial dimension and (macroscopic) shape of the contacts with the reservoirs. For the GEP our numerical simulations are in agreement with the predictions from the AP combined with the MFT \cite{EPL2013}. In one and two dimensions agreement is found for the first three cumulants. In three dimensions the first two cumulants agree with the AP, while the statistics for the third cumulant are insufficient for a reliable comparison. The diffusion coefficient, and as a result the current statistics, depends on the dimension. Only for the SSEP and the ZRP is the diffusion coefficient independent of the dimension.

A more precise numerical determination of the diffusion coefficient from Fick's first law is computationally very time consuming, at least using the methods presented here. It would therefore be of interest to find exact analytical results for the diffusion coefficient for the GEP. Another interesting question concerns the simulation of higher moments of the current distribution. This could be achieved using a sophisticated Monte Carlo algorithm to simulate rare events, see e.g.~\cite{PRL_giardina,lecomte2007numerical,PRL_takahiro}. Both the SSEP and the ZRP satisfy the sufficiency condition for the validity of the AP \eref{eq::sufficient}. However, many GEPs do not satisfy \eref{eq::sufficient}. Hence, one might observe deviations from the predictions of the AP for large current fluctuations. An analysis of the optimal density profiles, before and (possibly) after the dynamical phase transition, is also of interest.

The quantities $D(\rho)$ \cite{Naturekarger2014} and $\sigma(\rho)$ \cite{jobic2006influence} are experimentally accessible in nanoporous materials. The average particle flux through a system in contact with two particle reservoirs can also be measured \cite{book_diffnano}. If it is possible to measure the variance of the particle flux with a good precision, these techniques present an opportunity for an experimental verification of the additivity principle and, therefore, the macroscopic fluctuation theory.

\ack

We thank Christian Van den Broeck for bringing this problem to our attention. We are grateful to Bart Partoens and Carlo Vanderzande for a careful reading of the manuscript. This work was supported by the Flemish Science Foundation (Fonds Wetenschappelijk Onderzoek), Project No.~G038811N. The computational resources and services used in this work were provided by the VSC (Flemish Supercomputer Center), funded by the Hercules Foundation and the Flemish Government -- department EWI.

\appendix

\section{Algorithms\label{app::algorithm}}

Because we consider $\rho_A = 1$ and $\rho_B = 0$ for the SSEP, all transition rates are equal to one (also at the boundaries). All $n$ possible transitions are stored in a list. A random integer between $0$ and $n-1$ decides which transition takes place. The time between two events is taken from the distribution $p(t) = n \exp(-n t)$.
For the GEP with $\nmax = 2$ there are 12 different rates (four in the system and four at the contact with each reservoir). Since this is a small number, one can use the algorithm described by Schulze \cite{schulze2002kinetic}. For a fixed number of Monte Carlo steps, the computation time of both algorithms is constant for different system sizes.

\section{Data analysis\label{app::datacomp}}

The current fluctuations are measured as follows. First the system is allowed to relax to its steady state, after which we put the time at 0. The net number of particles that have entered the system between time 0 and $t$ is denoted by $Q_{t, 1}$. The net number of particles that have entered between time $t$ and $2 t$ is denoted by $Q_{t, 2}$, and so on. In the simulations $Q_t$ is determined by measuring the particle current at the left and right boundary. One then has a list $\{ Q_{t} \}$ with $N_l$ elements. The average is equal to 
\begin{equation}
\overline{Q_t} = \sum_{i=1}^{N_l} Q_{t, i} / N_l.
\end{equation}
For large $N_l$ the average $\overline{Q_t}$ is a good approximation for the average $\langle Q_t \rangle$ over $P(Q_t)$. The sample variance is equal to
\begin{equation}
S_t^2 = \sum_{i=1}^{N_l} (Q_{t, i} - \overline{Q_t})^2 / (N_l-1).
\end{equation}
For large $N_l$, $S_t^2$ converges to $\langle Q_t^2 \rangle - \langle Q_t \rangle^2$. 

The one-sigma error bar on $\overline{ Q_t }$ is equal to (assuming the $Q_{t,i}$'s are independent identically distributed variables)
\begin{equation}\label{eq::sigma}
\sigma = \sqrt{ S_t^2 / N_l}.
\end{equation}
The variance of $S_t^2$ is equal to
\begin{equation}
\mathrm{Var}(S_t^2) = \frac{1}{N_l} \left( \sigma_4 - \frac{N_l-3}{N_l-1} \sigma^4 \right),
\end{equation}
with $\sigma_4 = \langle (Q_t - \langle Q_t \rangle)^4 \rangle$ the fourth central moment of $P(Q_t)$ (see for example exercise 7.45 in \cite{casella2002statistical}). We estimate $\sigma$ by \eref{eq::sigma}. We do not estimate $\sigma_4$ directly from the simulation data, because our data do not allow for an accurate prediction of the fourth moment. Rather, we use the prediction for $\sigma_4$ from the AP \cite{bodineau2004}. One-sigma error bars on $S_t^2$ are equal to $\left[\mathrm{Var}\left(S_t^2 \right)\right]^{1/2}$. Except for the third cumulant, all other error bars are obtained from addition and multiplication of $\overline{Q_t}$ and $S_t^2$. The rules for finding these error bars can be found in e.g.~\cite{error_book}. The Fano factor is calculated by $F(t) = S_t^2 / \overline{Q_t}$. The error bar on the third cumulant is found by bootstrapping the simulated data.

By adding the currents pairwise $Q_{t, i} + Q_{t, i+1}$ (with $i$ odd), one can calculate $\overline{Q_{2 t}}$ and $S_{2t}^2$ for the time interval $2 t$ (with $N_l / 2$ points), and so on. We study the Fano factor $F(nt)$ for $1 \leq n \leq 6$.

\begin{figure}%
\subfloat[][]{
\includegraphics[width=0.5\columnwidth]{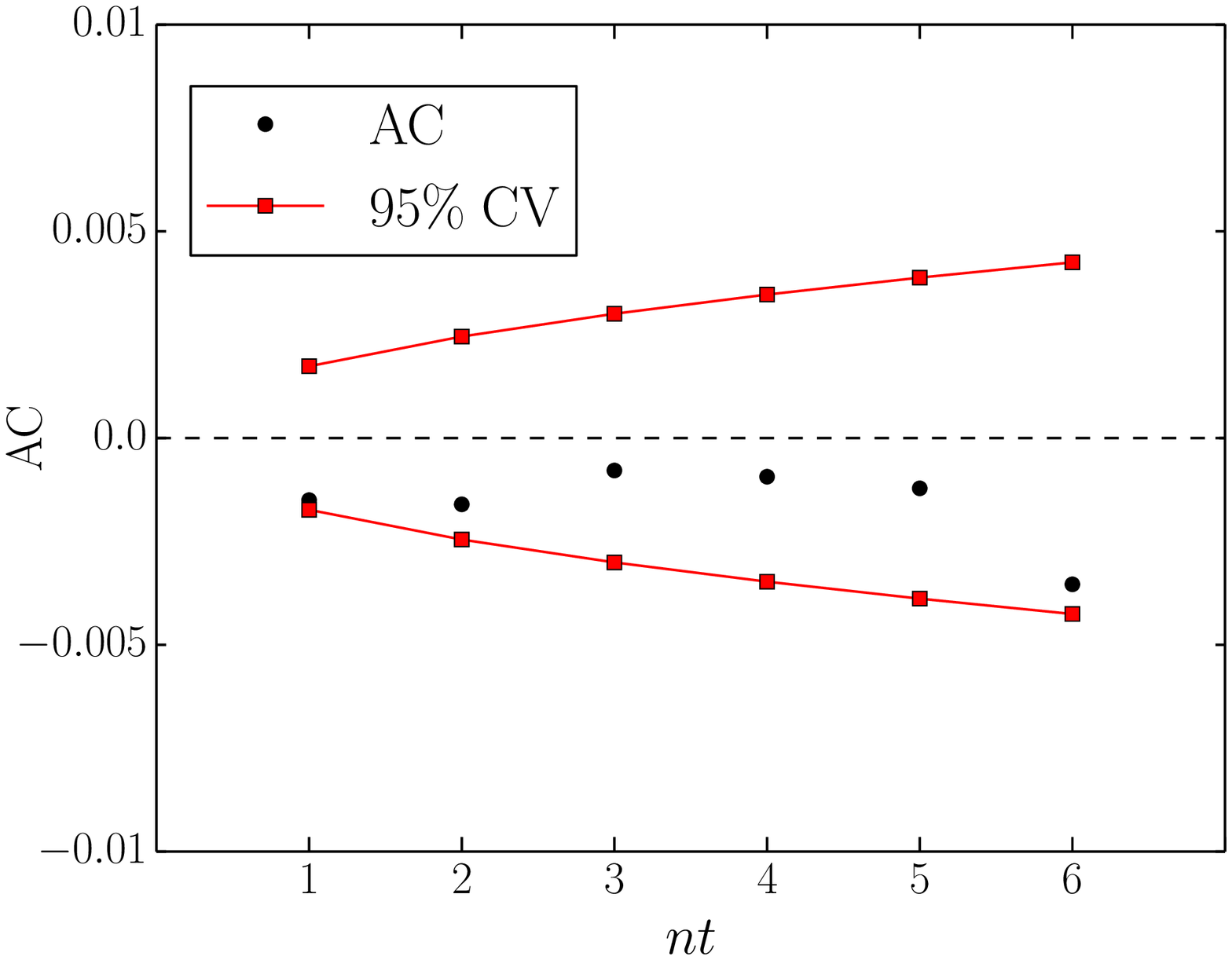}
\label{fig::AC}
}%
\subfloat[][]{
\includegraphics[width=0.5\columnwidth]{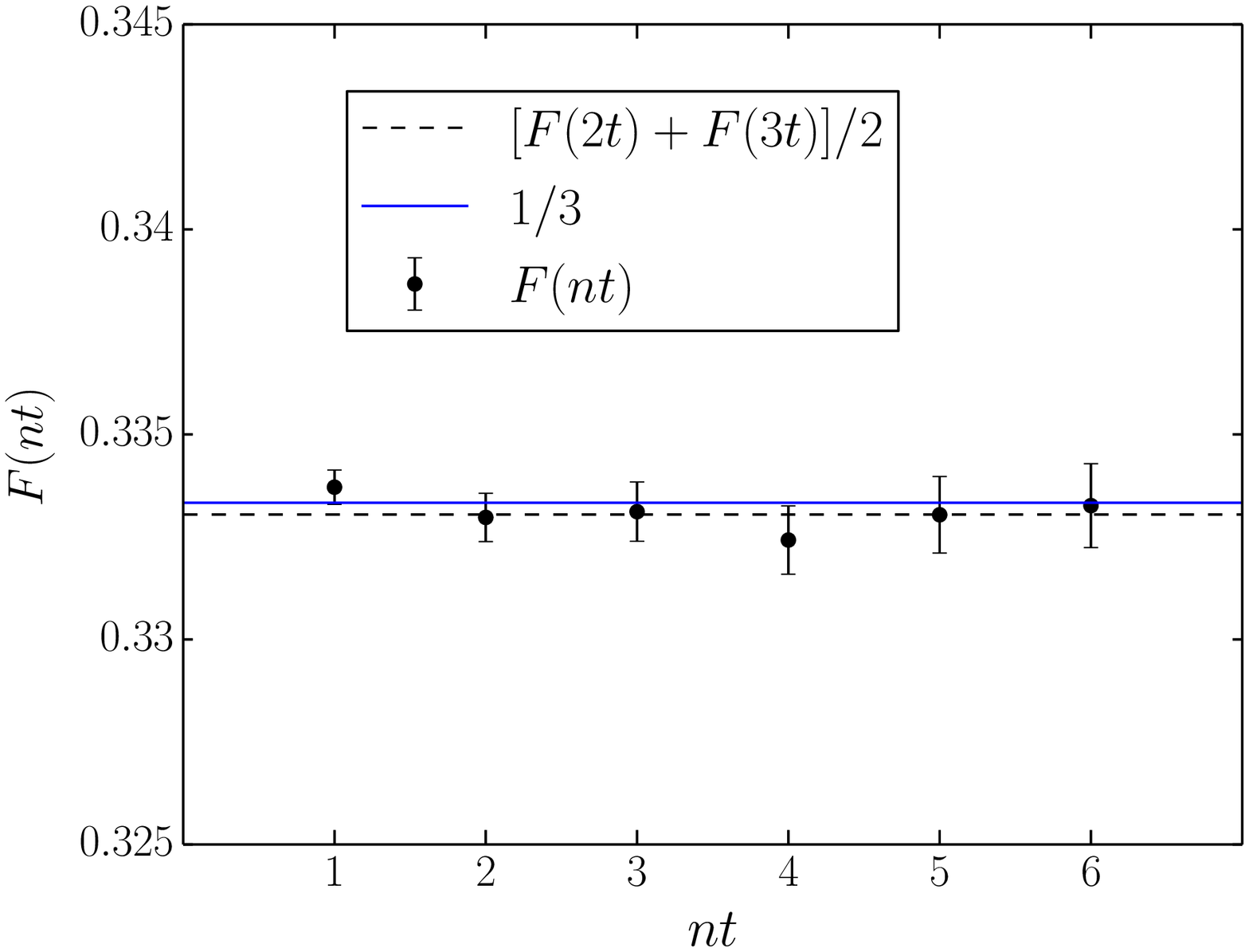}
\label{fig::fano_Qt}
}
\caption{The two-dimensional SSEP with $t = 2.10^4$, $L = 40$, and the geometry of Figure \ref{fig::SSEP_lattice_2D}. (a) (circles) Autocorrelation \eref{eq::ACdef}. (red squares) Critical values to reject the null hypothesis AC = 0 at 95 $\%$ significance level. (b) (circles) $F(nt)$. (dashed line) Average of $F(2t)$ and $F(3t)$. This is the value of the data point in Figure \ref{fig::fano_SSEP}.}%
\label{fig:cont}%
\end{figure}

We now explain how we check if the data have converged in time. For clarity we consider the specific example of the two-dimensional SSEP at $L = 40$ with $t = 2.10^4$. 
The autocorrelation (AC) of $Q_{t, i}$ and $Q_{t, i+1}$ is
\begin{equation}\label{eq::ACdef}
\mathrm{AC} = \frac{\sum_{i=1}^{N_l-1} (Q_{t, i} - \overline{ Q_t })(Q_{t, i+1} - \overline{ Q_t }) }{\sum_{i=1}^{N_l} (Q_{t, i} - \overline{ Q_t })^2}.
\end{equation}
The AC is plotted in Figure \ref{fig::AC}, together with the critical values (CVs) to reject the null hypothesis that AC = 0 at 95 $\%$ significance level. All points are smaller than the CVs. The point at $n=1$ is, however, very close to the lower CV. This suggests that there is still a non-negligible AC for times $1t$. Indeed, for small times the AC is always negative. For large times, when the $Q_{t,i}$'s are uncorrelated, the AC fluctuates close to zero. The scale of ``close to zero'' is determined by the CVs.

\begin{figure}
\centering
\includegraphics[width=0.5\columnwidth]{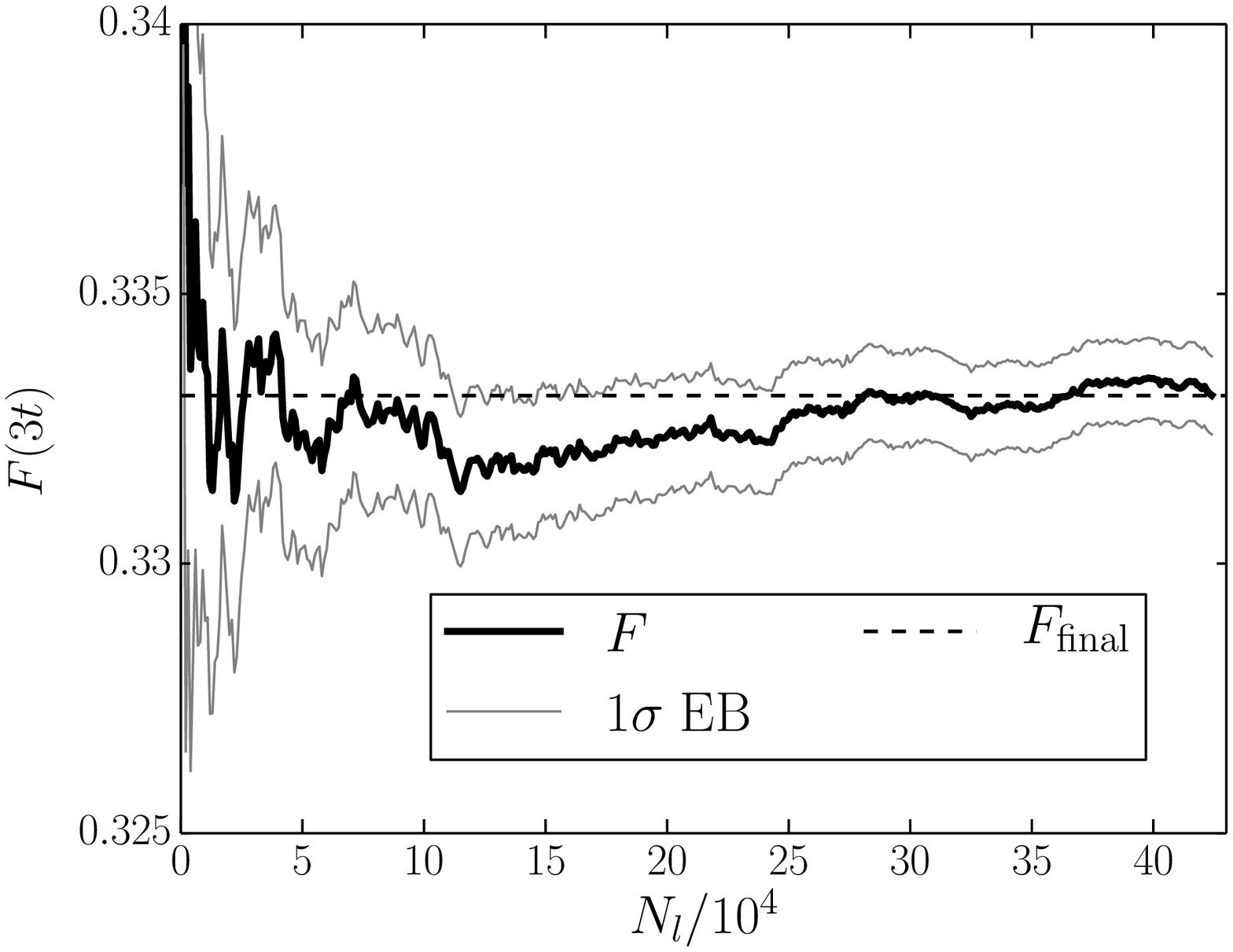}
\caption{The two-dimensional SSEP with $t = 2.10^4$, $L = 40$, and the geometry of Figure \ref{fig::SSEP_lattice_2D}. (thick black line) $F(3t)$ after $N_l$ simulated points. (thin grey lines) one-sigma error bars. (dashed line) final value of $F(3t)$.}
\label{fig::checkconv}
\end{figure}

The Fano factor $F(n t)$ is plotted in Figure \ref{fig::fano_Qt}. $F(1t)$ is slightly higher than the other 5 points, indicating again that there is not yet convergence in time. The first two point that are converged in time are $F(2t)$ and $F(3t)$. A plot as a function of the number of simulated points $N_l$ for $F(3t)$ is shown in Figure \ref{fig::checkconv}. After $N_l \approx 25 . 10^{4}$ the data fluctuate around the end value $F_{\mathrm{final}}$, indicating a good convergence for $F(3t)$.
The average of $F(2t)$ and $F(3t)$ is taken as the final data point (as plotted in Figure \ref{fig::fano_SSEP}). For most points, the first two converged values are averaged to calculate the final result. If computation times are exceedingly long, such as for the SSEP in two dimensions for $L=50$, only the first converged point is taken. In this case this point is $F(2t)$. $F(3t)$ has not yet converged as can be seen from a graph similar to Figure \ref{fig::checkconv}. This explains the large error bar for $L = 50$ compared to the other points for the two-dimensional SSEP.

\section{Simulation of diffusion coefficient\label{app::simD}}

$D(\rho)$ is simulated for 30 concentrations, see \cite{PRL2013,PREbecker} for details on the simulations and calculation of the error bars. In this paper the length in the $x$ direction is $L = N+1 = 16$ in two and three dimensions. In one dimension the analysis was performed for $L=21$ and $L=16$. The predicted values of $I_1$ were the same up to a relative difference of $0.006 \%$. The data in the paper are for $L=21$ in one dimension. The concentration gradient for low and high concentrations is taken between $\Delta \rho = 0.05$ and $\Delta \rho = 0.03$. For the other concentrations we take $\Delta \rho = 0.06$. The values at $\rho = 0$ and $\rho = \nmax$ can be calculated analytically: $D(0) = 1$ and $D(2) = 2$. An approximation for the continuous function $D(\rho)$ is achieved by interpolating these 32 points (using the ``Interpolation'' function of Mathematica). For concentrations smaller than $\rho \approx 0.04$ and higher than $\rho \approx 1.96$ the interpolated values are higher than the uncorrelated result \eref{eq::DtDMF}. Since we know that correlations lower $D(\rho)$, we consider the uncorrelated results for these concentrations instead of the interpolated function.

\section{Cumulant generating function in $d>1$\label{app::Cd}}

The CGF $\mu_d(\lambda)$ of a $d$-dimensional system is equal to (cf.~the last equation in \cite{EPL2013})
\begin{equation}
\mu_d(\lambda) = \left[ L^{d-2} \int d \vec{r} \left( \vec{\nabla} v(\vec{r})  \right)^2  \right] \times \left[ L \mu_1(\lambda) \right].
\end{equation}
$\mu_1(\lambda)$ is the CGF of a one-dimensional system described by $D_d(\rho)$ and $\sigma_d(\rho)$.
Consider a rectangular system of length $L_x$ and height $L_y$. All sites at $x = 0$ are coupled to reservoir $A$ and all sites at $x=L_x$ are coupled to reservoir $B$. 
$L$ is the typical domain size, which we take equal to $L_x$.
$v(x,y)$ is a function on the domain $0 \leq x \leq 1$, $0 \leq y \leq L_y/L_x$, that satisfies the Laplace equation $\Delta v(\vec{r}) = 0$, with $v(0,y) = 0$, $v(1,y) = 1$, and Neumann boundary conditions otherwise. For the geometry we consider it is straightforward to show that $v(x,y) = x$. One then finds
\begin{equation}
\mu_2 (\lambda) = \left[ \int_0^1 \int_0^{L_y/L_x} dx dy \right] \times \left[ L_x \mu_1(\lambda) \right] = L_y \mu_1(\lambda).
\end{equation}
The calculation for the same geometry in three dimensions shows that $\mu_3 (\lambda) = L_y L_z \mu_1(\lambda)$.

The density $\rho(x,y)$ can be found from the one-dimensional profile $\rho_1(x)$ (equation (33) in \cite{EPL2013})
\begin{equation}\label{eq::rho12}
\rho(x,y) = \rho_1(v(x,y)) = \rho_1(x).
\end{equation}
Note that the only assumption required for these results is the time-independence of the optimal density and current profiles.
In the study of the two-dimensional KMP model with all the boundary sites connected to reservoirs \cite{JSP_Hurtado,pe2011large_article}, one made the extra assumption that the optimal current profile is constant $\vec{j}_{\vec{J}}(\vec{r}) = \vec{J}$. This extra assumption is unnecessary: it can be derived from the time-independence of the optimal profiles and the MFT. Indeed, in one dimension time-independent profiles imply a constant current profile. A constant current profile in two dimensions follows from \eref{eq::rho12}. Note that for more general couplings to the reservoirs, such as in Figure \ref{fig::SSEP_lattice}, the optimal current profile need not be constant.

We have solved numerically the Laplace equation for $v(\vec{r})$ for the domain in Figure \ref{fig::fano_SSEP}. One finds $\mu_2(\lambda) \approx 0.663 L \mu_1(\lambda)$. This agrees with our kMC results, as discussed in Section \ref{sec::SSEP}.
 

\providecommand{\newblock}{}

\end{document}